\pdfoutput=1

\documentclass[11pt]{article}

\usepackage[preprint]{acl}

\usepackage{times}
\usepackage{latexsym}

\usepackage[T1]{fontenc}

\usepackage[utf8]{inputenc}

\usepackage{microtype}

\usepackage{inconsolata}

\usepackage{graphicx}
\usepackage{subfigure}
\usepackage{booktabs} 
\usepackage{mdframed}
\usepackage{colortbl}
\usepackage{multirow}
\usepackage{float}
\usepackage{pifont}
\usepackage{diagbox}
\usepackage{url}

\usepackage{hyperref}
\usepackage{amsmath}
\usepackage{amssymb}
\usepackage{mathtools}
\usepackage{amsthm}
\usepackage{array}
\usepackage{multirow}
\usepackage[capitalize,noabbrev]{cleveref}
\definecolor{lightergray}{gray}{0.90}
\newcolumntype{a}{>{\columncolor{lightergray}}c}
\newcolumntype{x}{>{\columncolor{lightergray}}l}
\newcommand{\dataname}{\textsc{RealEdit}}
\newcommand{\modelname}{\textsc{VoiceCraft}}
\newcolumntype{L}[1]{>{\arraybackslash}p{#1}}

\usepackage{xcolor}

\newcommand{\pyp}[1]{{\color{purple}[Puyuan: #1]}}
\newcommand{\david}[1]{{\color{brown}[David: #1]}}
\renewcommand{\david}[1]{}
\renewcommand{\pyp}[1]{}

\title{\modelname: Zero-Shot Speech Editing and Text-to-Speech in the Wild}

\author{Puyuan Peng$^1$ \qquad Po-Yao Huang$^2$ \qquad Shang-Wen Li$^2$ \\ \textbf{Abdelrahman Mohamed}$^3$ \qquad \textbf{David Harwath}$^1$ \\
        $^1$The University of Texas at Austin \qquad $^2$FAIR, Meta \qquad $^3$Rembrand\\
\texttt{pyp@utexas.edu}}

\begin{document}
\maketitle
\begin{abstract}
    We introduce \modelname, a token infilling neural codec language model, that achieves state-of-the-art performance on both speech editing and zero-shot text-to-speech (TTS) on audiobooks, internet videos, and podcasts\footnote{Data, code, and model weights are available at \url{https://github.com/jasonppy/VoiceCraft}.}.
    \modelname~employs a Transformer decoder architecture and introduces a token rearrangement procedure that combines causal masking and delayed stacking to enable generation within an existing sequence.
    On speech editing tasks, \modelname~produces edited speech that is nearly indistinguishable from unedited recordings in terms of naturalness, as evaluated by humans; 
    for zero-shot TTS, our model outperforms prior SotA models including VALL-E and the popular commercial model XTTS v2. 
    Crucially, the models are evaluated on challenging and realistic datasets, that consist of diverse accents, speaking styles, recording conditions, and background noise and music, and our model performs consistently well compared to other models and real recordings. 
    In particular, for speech editing evaluation, we introduce a high quality, challenging, and realistic dataset named \dataname. 
    We encourage readers to listen to the demos at~\url{https://jasonppy.github.io/VoiceCraft_web}.
\end{abstract}

\section{Introduction}
\begin{figure}
    \centering
    \includegraphics[width=\linewidth]{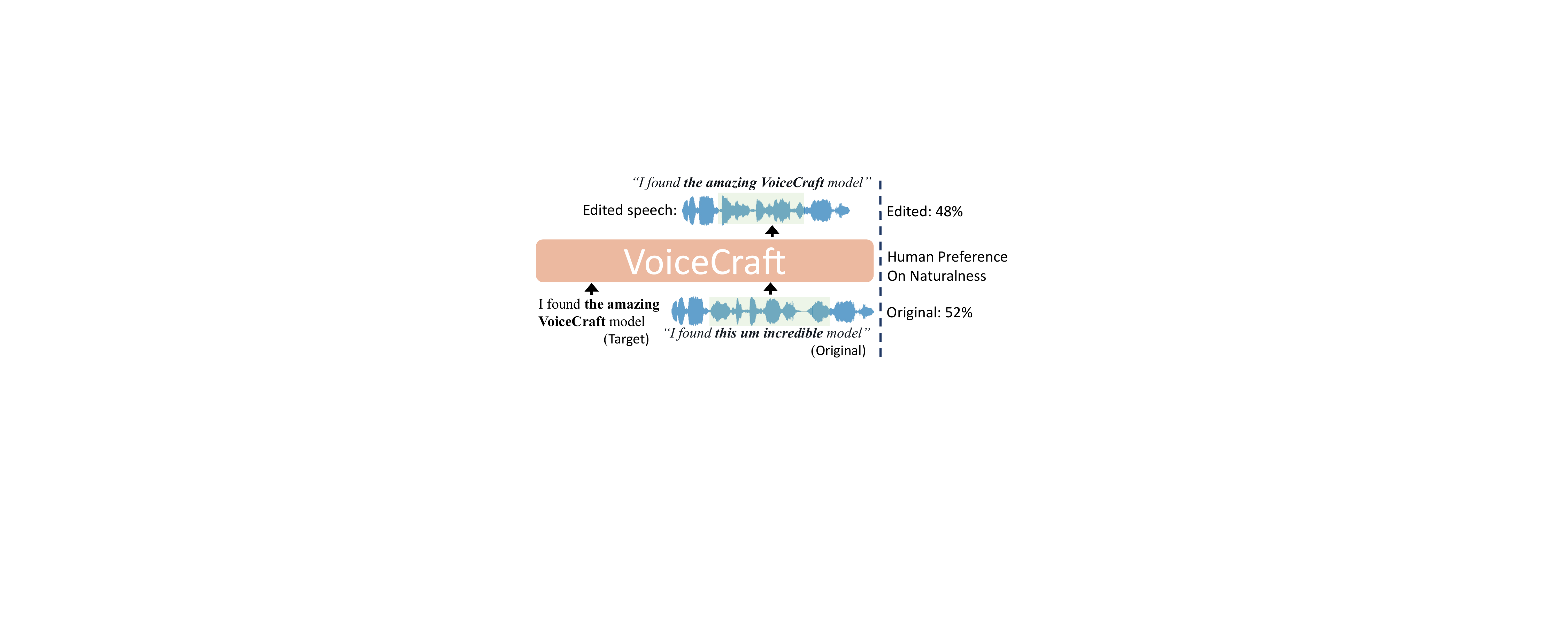}
    \caption{Speech editing with \modelname. Human listeners prefer \modelname~edited speech over the original real recording 48\% of the time in side-by-side naturalness comparison (details in \S\ref{sec:se})}\label{fig:teaser}
    \vspace{-1em}
\end{figure}
We introduce \modelname, a Transformer-based neural codec language model (NCLM) that performs infilling generation of neural speech codec tokens autoregressively conditioned on bidirectional context. \modelname~achieves state-of-the-art (SotA) performance on both speech editing (shown in Fig.~\ref{fig:teaser}) and zero-shot TTS. 
Our method is based on a two-step token rearrangement procedure that consists of a \emph{causal masking} step and \emph{delayed stacking} step.
The causal masking technique is inspired by the success of causal masked multimodal model in joint text-image modeling~\citep{cm3}, and our proposed technique works for speech codec sequences, which enables autoregressive generation with bidirectional context. 
In addition, we further integrate causal masking with delayed stacking~\citep{delayed,musicgen} as our proposed token rearrangement procedure, to ensure efficient multi-codebook modeling. 

To evaluate speech editing, we manually crafted a first-of-its-kind, realistic, and challenging dataset named \dataname. 
\dataname~consists of $310$ real world speech editing examples, with waveforms sourced from audiobooks~\citep{libritts}, YouTube videos~\citep{GigaSpeech2021}, and Spotify podcasts~\citep{clifton-etal-2020-100000}, and duration ranging from $5$ seconds to $12$ seconds. 
To create the target transcripts, the transcripts of the source speech are edited in such a way that the edited transcripts remain grammatically correct and are semantically coherent. 
The dataset is designed to cover a wide range of editing scenarios, including insertion, deletion, substitution, and multi-span editing, with the length of the edited text ranging from $1$ word to $16$ words. 
Compared to commonly used speech synthesis evaluation datasets that only contain audiobooks such as VCTK~\citep{Yamagishi2019CSTRVC}, LJSpeech~\citep{ljspeech17}, and LibriTTS~\cite{libritts}, \dataname~is more challenging in that the recordings have diverse content, accents, speaking styles, recording conditions, and background sounds. We believe that the realism and diversity of \dataname~makes it a reliable indicator of the practicality of speech editing models in the real world.

In the subjective human listening tests, \modelname~significantly outperforms prior SotA speech editing model on \dataname.
Importantly, the edited speech produced by \modelname~is nearly indistinguishable from the original unedited recording in terms of naturalness. 
We found that \modelname~ generalizes well to zero-shot TTS without any finetuning, achieving SotA performance on a dataset comprised of audiobooks and YouTube videos, outperforming strong baselines including reproduced VALL-E~\cite{Wang2023NeuralCL} and the popular commercial model XTTS v2~\cite{xttsv2}.
In summary, our contributions are: 
\begin{enumerate}
    \item We introduce \modelname, a neural codec language model for speech editing that generates synthesized speech that is nearly indistinguishable from in-the-wild recordings according to human listeners. We also release the code and model weights for \modelname.
    \item We show that \modelname~generalizes well to zero-shot TTS without finetuning.
    \item We release a high quality, challenging, and realistic speech editing evaluation dataset \dataname.
\end{enumerate}
 
\section{Related Work}
\textbf{Neural codec langauge models (NCLM) and zero-shot TTS.}
Tokenizing speech signals into sequences of learnable, discrete units and then training a language model on the resulting unit sequences was initially proposed in the context of textless NLP~\citep{hsu_harwath_glass_2021,Lakhotia2021OnGS,Kharitonov2021TextFreePG,Nguyen2022GenerativeSD}, where the goal is to perform NLP tasks directly on spoken utterances without the need to first transcribe the speech into text. 
Recently, NCLMs that operates on tokens from Residual vector quantization (RVQ)-based models~\citep{Zeghidour2021SoundStreamAE,Defossez2022HighFN} attract increased attention due to its high quality generation. For example, AudioLM~\citep{Borsos2022AudioLMAL} exhibits strong performance on long-term coherent speech continuation. 
Zero-shot TTS is a task where a model needs to synthesize speech in a target voice which was unseen during training, given only the target transcript and a short reference recording of the target voice. Framing zero-shot TTS as transcript-conditioned speech continuation, VALL-E~\citep{Wang2023NeuralCL} and Spear-TTS~\cite{Kharitonov2023SpeakRA} are the first applications of NCLMs on this task, significantly outperforming non-NCLM approaches. 
\citet{Zhang2023SpeakFL} extends VALL-E to cross-lingual TTS. \citet{Guo2022PromptttsCT,Yang2023InstructTTSME,Liu2023PromptStyleCS,Ji2023TextrolSpeechAT,Lyth2024NaturalLG} adapt NCLMs style-controlled speech synthesis. \citet{Song2024ELLAVSN,Du2024VALLTDG} enhance phoneme alignment in NCLMs to reduce error. \citet{Wang2023VioLAUC} proposes a unified NCLM for both speech generation and recognition tasks. \citet{Borsos2023SoundStormEP} proposes an efficient parallel decoding method. \citet{jiang2024megatts} proposes disentangled timbre and prosody modeling, where the latter is modeled with a NCLM. NCLMs have also been successfully applied to other audio domains.  \citet{Kreuk2022AudioGenTG} applies NCLM to sound effects generation, and \citet{Agostinelli2023MusicLMGM,Donahue2023SingSongGM,Garcia2023VampNetMG,musicgen} use NCLMs for music generation.

\textbf{Speech editing.} This task requires a model to alter words or phrases within an utterance to match a target transcript, but the regions of the original speech not targeted for editing must remain unchanged (see Fig.~\ref{fig:teaser} for an example). Early methods achieve text-guided speech insertion and substitution by combining a single speaker TTS model and a voice conversion model to generate desired speech segment, which is then concatenated with unedited part~\cite{Jin2017VoCoTI}. Since the generation is not conditioned on the unedited part of the speech, the result sounds unnatural due to prosody mismatch and boundary artifacts~\cite{Morrison2021ContextAwarePC}. More recent speech editing models have attempted to condition their generation on surrounding speech context. \citet{Tan2021EditSpeechAT} uses two unidirectional LSTM models with bidirectional fusion. \citet{Wang2022ContextAwareMP,Bai2022A3TAA,Borsos2022SpeechPainterTS} uses the masked reconstruction objective with Convolutional or Transformer models to further improve contextualization. FluentSpeech~\citep{Jiang2023FluentSpeechSA} is a  diffusion-based speech editing model that achieves SotA performance on speech editing on LibriTTS and VCTK.

The research community starts to investigate the possibility of having a unified model for both zero-shot TTS and speech editing. \citet{Yin2022RetrieverTTSMD,Jiang2023MegaTTSZT} propose modular models for the two tasks, while our model is end-to-end. Concurrent work SpeechX~\citep{Wang2023SpeechXNC} adapt VALL-E by prompt tuning for a range of tasks including speech editing and zero-shot TTS, but no human evaluation is conducted in their paper. Concurrent work UniCATS~\cite{UniCATS} is a diffusion-based modular model for the two tasks. 
However their model is only evaluated on masked speech reconstruction of span length less than 2 seconds, while our model is evaluated on as much as 16 words editing. Voicebox~\citep{Le2023VoiceboxTM} is a recent flow matching based model capable of a wide range of tasks including speech editing and zero-shot TTS. 
However the speech editing capability is not evaluated in their paper, and only shown in their demo page. 
We therefore compare our model's editing results with Voicebox's on our demo page using on the same examples from their demo page.
\section{Method} 
\modelname~casts both sequence infilling (for speech editing) and  continuation (for zero-shot TTS) as a simple left-to-right language modeling by rearranging neural codec's output tokens. 
The rearrangement involves two steps: (1) causal masking (\S\ref{ssec:causal_masking}) to enable autoregressive continuation/infilling with bidirectional context  
and (2) delayed stacking (\S\ref{ssec:delayed_stacking}) to ensure efficient multi-codebook modeling.
\modelname~employs decoder-only Transformers and is trained with an autoregressive sequence prediction (\S\ref{ssec:modeling}).
We introduce the inference setup for speech editing and zero-shot TTS in \S\ref{ssec:inference}.

\begin{figure*}
    \centering
    \includegraphics[width=\textwidth]{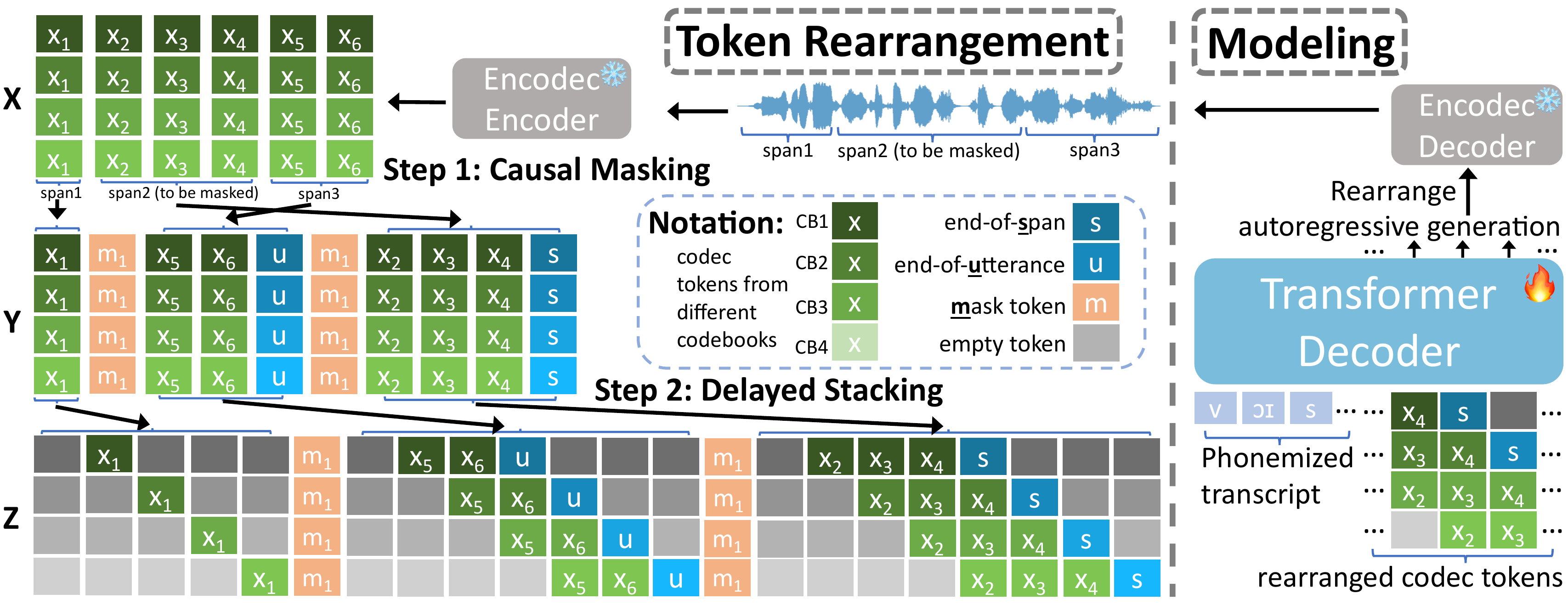}
    \vspace{-0.6cm}
    \caption{An example of the token rearrangement procedure and modeling framework. The rearrangement procedure involves two steps: (1) \emph{Causal masking}, where masked spans are replaced with mask tokens and moved to the end, and (2) \emph{Delayed stacking}, where tokens are shifted in the time dimension based on their codebook index.}
    \label{fig:model}
    \vspace{-0.3cm}
\end{figure*}
\subsection{Rearrangement Step 1: Causal Masking}
\label{ssec:causal_masking}
As shown on the left hand side of Fig.~\ref{fig:model}, given a continuous speech waveform as input, we first use Encodec~\citep{Defossez2022HighFN} to quantize it into a $T$ by $K$ codec matrix $X$, where $T$ is the number of temporal frames, and $K$ is the number of RVQ codebooks. 
$X$ can be written as $(X_1, \cdots, X_T)$, where $X_t$ is a vector of length $K$ representing the codes from different codebooks at time step $t$, and we assume that code from codebook $k$ models the residual from codebook $k-1$. During training, our goal is to randomly mask some span of tokens $(X_{t_0}, \dots, X_{t_1})$, and then autoregressively predict these masked tokens conditioned on all of the unmasked tokens. This is a problem when $t_1 < T$, because we cannot condition on future outputs when performing autoregressive generation. We need to modify the masking on $X$ so that it is \textit{causal}, by moving the span to be masked to the end of the sequence, so that when infilling these tokens the model can condition on both past and future unmasked tokens~\cite{cm3,donahue-etal-2020-enabling,Bavarian2022EfficientTO}. 

The procedure outlined above can be trivially extended to multiple masked spans by simply moving \textit{all} masked spans to the end of the sequence. The number of spans to be masked $n$ is sampled from $\text{Poison}(\lambda)$, and then for each span, we sample a span length $l \sim \text{Uniform}(1,L)$. Finally, we randomly select the locations of the spans within $X$ under the constraint that they do not overlap with each other. The selected $n$ spans are then replaced with mask tokens $\text{\textlangle M$_1$\textrangle}, \cdots, \text{\textlangle M$_n$\textrangle}$. The original tokens within these masked spans are moved to the end of the sequence $X$, with each span preceded by its corresponding mask token. 

Consider this example: let $X=(X_1, \dots, X_6)$ and imagine we wish to mask a single span from $X_2$ to $X_4$. The original sequence $X$ is rearranged into $Y=(Y_1; \text{\textlangle M$_1$\textrangle}; Y_2;\text{\textlangle M$_1$\textrangle}; Y_3;)$, where $Y_1=(X_1)$, $Y_2 = (X_5, X_6)$, and $Y_3 = (X_2, X_3, X_4)$. We call $Y_1$ and $Y_2$ the unmasked spans, and $Y_3$ the masked span. An \textit{end of span} or \texttt{EOS} token is added to the end of each masked span (in this example at the end of $Y_3$), and an \textit{end of utterance} or \texttt{EOU} token is added to the end of the utterance (i.e. $Y_2$). For simplicity, we do not explicitly denote these special tokens and assume they are part of the spans.

\subsection{Rearrangement Step 2: Delayed Stacking}
\label{ssec:delayed_stacking}
After the causal masking token rearrangement, each timestep of the rearranged matrix $Y$ is vector of $K$ tokens. \citet{musicgen} observed that when performing autoregressive generation over stacked RVQ tokens, it is advantageous to apply a \textit{delay pattern} so that the prediction of codebook $k$ at time $t$ can be conditioned on the prediction of codebook $k-1$ from the same timestep. We take a similar approach which we describe here.
Assume a span $Y_s$ is of shape $L_s \times K$. Applying the delay pattern rearranges it into $Z_s = (Z_{s,0}, Z_{s,1}, \cdots,Z_{s,L_s + K -1})$, where $Z_{s,t}, t\in[L_s+K-1]$ is defined as\footnote{$[N]$ represents integer set $\{0,1,\cdots,N\}$}:
\begin{align}
    Z_{s,t} = (Y_{s,t,1}, Y_{s,t+1,2}, \cdots, Y_{s,t-K+1,K})
\end{align}
where $Y_{s, t-k+1, k}$ denotes the token located at coordinate $(t-k+1, k)$ in matrix $Y_s$, i.e. the $k$th codebook entry at the $(t-k+1)$th timestep. To make sure that $\forall t \in [L_s+K-1] $, $Z_{s,t}$ contains $K$ valid tokens, we introduce a special learnable $\texttt{[empty]}$ token and define $Y_{s, t-k+1, k} \triangleq \texttt{[empty]} \,, \forall t\in\{s:s<k \cup s-k+1>L_s\}$.
Note that the mask tokens are not part of any span and are not changed during delayed stacking. We define the resulting matrix of delayed stacking $Z = (Z_1, \text{\textlangle M$_1$\textrangle}, Z_2, \text{\textlangle M$_1$\textrangle},\cdots,\text{\textlangle M$_{\frac{S-1}{2}}$\textrangle}, Z_S)$ (assuming $Y$ consists of $S$ spans). See the diagram for $Z$ in Fig.~\ref{fig:model} for an illustration. 

\subsection{Modeling}
\label{ssec:modeling}
As shown in the right hand side of Fig.~\ref{fig:model}, we use a Transformer decoder to model $Z$ autoregressively, conditioned on transcript of the speech $W$. Therefore, the input to the decoder is $[W; Z]$, where ``;'' denotes concatenation. At timestep $t$ of span $s$ in codec matrix $Z$, the model predicts all $K$ tokens of $Z_{s,t}$ simultaneously, by using $K$ MLP heads to project the transformer's final hidden state to $K$ sets of logits, one for each of the $K$ codebooks. Note that the prediction is conditioned on transcript $W$, and all tokens in $Z$ before $Z_{s,t}$, denoted as $H_{s,t}$. Mathematically, the Transformer decoder models the factorized conditional distribution of $Z$:
\begin{align}
    \mathbb{P}_{\theta}(Z|W) &= \prod_{s}\prod_{t} \mathbb{P}_{\theta}(Z_{s,t}|W, H_{s,t}) \label{eq:autoregressive}\\
           &= \prod_{s}\prod_{t} \prod_{k=1}^K \mathbb{P}_{\theta}(Z_{s,t,k}|W, H_{s,t}) \label{eq:factorize} 
\end{align}
Where $\theta$ represent the parameters of the model. Equation~\ref{eq:autoregressive} is the autoregressive factorization across time, while Equation~\ref{eq:factorize} is the factorization across codebooks given an independence assumption - given $W$ and $H_{s,t}$, the $K$ RVQ codes in $Z_{s,t}$ are assumed to be independent of each other. We argue in appendix~\ref{sec:independence} that this assumption is mild.

With the token level probability formulation in Equation~\ref{eq:factorize}, we derive the training loss as the negative log likelihood $\mathcal{L}(\theta) = -\log \mathbb{P}_{\theta}(Z|W) = -\sum_{k=1}^K \mathcal{L}_k(\theta)$. Empirically, we found that weighting the first residual codebooks more than the latter codebooks leads to better performance, and therefore our final loss is $\mathcal{L}(\theta) = \sum_{k=1}^K \alpha_k \mathcal{L}_k(\theta)$, where $(\alpha_k)_{k=1}^K$ are tunable hyperparameters. Note that we follow~\citet{cm3} and calculate the prediction loss on all tokens (not just the tokens in the masked spans), except for mask tokens and \texttt{[empty]} tokens.

\begin{table*}
    \caption{Examples of the speech editing dataset \dataname. More examples are shown in table~\ref{tab:se_examples_app}.}\label{tab:se_examples}
    \vspace{-0.3cm}
    \centering
    \resizebox{\textwidth}{!}{%
    \begin{tabular}{L{0.13\textwidth}L{0.5\textwidth}L{0.5\textwidth}}
        \toprule
        Edit Types & Original & Edited \\
        \midrule
        deletion & I wrote the title \textbf{of the course many years ago, ah,} when I created this course. & I wrote the \textbf{title when} I created this course. \\
        \midrule
        insertion & And \textbf{we're at} this point. & And we're \textbf{all extremely excited} at this point. \\
        \midrule
        substitution, substitution & See why it's extremely \textbf{valuable to it's kind of like} it's kind of like having a \textbf{wall hack} to watch a demo. & See why it's extremely \textbf{important right?} it's kind of like having a \textbf{rough time} to watch a demo. \\
        \bottomrule
    \end{tabular}
    }
    \vspace{-0.5cm}
\end{table*}

\subsection{Inference}\label{sec:inference}
\label{ssec:inference}
\textbf{Speech Editing.} 
The setting for speech editing is the following: we have a speech recording $R$ and its transcript $W$, and we want the model to modify only the relevant spans of $R$ so that it matches the target transcript $W'$. We assume that $W'$ is an edited version of $W$, where some words have been inserted, substituted, or deleted.
This task is almost exactly the same as the training task, with two differences: 
1) during training, the input transcript is simply the transcript of the original recording $W$, while during inference it is a modified transcript $W'$
2) during training, the spans to be masked (i.e. edited) are chosen randomly. During inference, we select them by comparing the original transcript and the target transcript to identify the words that should be masked out, and then use the word level forced alignment of the original transcript to identify the codec token spans that correspond to these words to be masked.
To ensure a smooth transition between the edited speech and the unedited speech, the neighboring words surrounding the span to be edited also need to be slightly modified in order to model co-articulation effects. Therefore, we specify a small margin hyperparameter $\epsilon$, and extend the mask span length by $\epsilon$ on both the left and right sides\footnote{for substitution and deletion, the spans that are to be masked are just those words that are different from the target plus the margin; for insertion, the spans are just left and right margin spanning from the middle of the two words where the insertion happens}. During autoregressive generation, we feed the model the target transcript with all unmasked spans, with mask tokens inserted in the locations where the edits should take place. We then have the model autoregressively continue this sequence, whereby it fills in the masked spans. The generated codec tokens are then spliced back into their correct location in the utterance, and we map the complete codec token sequence back to a waveform using the Encodec decoder network.

\textbf{Zero-shot TTS.} 
As we previously noted, zero-shot TTS for our model is straightforward because it simply corresponds to performing an insertion edit at the end of the original utterance. In this case, the model is provided a voice prompt with its transcription, as well as the target transcript of the speech to be generated. The three inputs are concatenated together and fed to the model, after which it generates the codec sequence of the target transcript autoregressively.

\section{\dataname: a realistic and challenging speech editing dataset}
To support as realistic an evaluation as possible, we constructed a \textbf{first-of-its-kind} dataset of $310$ manually-crafted speech editing examples. 
Each example consists of a tuple: (original audio, original transcript, edited transcript). The dataset contains $100$ utterances from LibriTTS (dev-clean and dev-other)~\citep{libritts}, $100$ utterances from YouTube (from Gigaspeech testset)~\citep{GigaSpeech2021} and $110$ utterances from the Spotify Podcast dataset~\citep{clifton-etal-2020-100000}. 
We manually checked the utterances for accuracy, then had native English speakers revise them to create edited transcripts. 
For each utterance, we determine the type of modification using predefined probability distributions of editing type, number of disjoint spans to be edited, and editing span length.
Specifically, we study the following categories: 1) number of edited spans: $1$ or $2$; 2) type of edits: \textit{insertion}, \textit{deletion} and \textit{substitution}; 3) editing span length: short ($1$-$2$ words), medium ($3$-$6$ words), long ($7$-$12$ words). Crucially, a edited transcript must be \textbf{grammatically correct and semantically coherent}. Examples of the dataset are shown in table~\ref{tab:se_examples} and~\ref{tab:se_examples_app}, and statistics are shown in table~\ref{tab:se_data}, 
\begin{table}[h]
    \small
    \caption{Dataset statistics for speech editing evaluation. Note that for 2-span editing, each example is edited using $2$ of the $3$ edit types.}\label{tab:se_data}
    \vspace{-0.4cm}
    \begin{center}
    \resizebox{\columnwidth}{!}{
    \begin{tabular}{l|ccc|c}
        \toprule
        \diagbox[dir=NW,width=9em,height=1.6em]{length}{type}& Insert. & Delet. & Substi. & Total \\
        \midrule
        1-2 words (1 span) & 8 & 17 & 38&63 \\
        3-6 words (1 span) & 22 & 24 & 79&125 \\
        7-12 words (1 span) & 15 & 11 & 56&82 \\
        \midrule
        1 span total &45 & 52 & 173&270 \\
        \midrule
        2 spans total & 13 & 13 & 54 &40 \\
        \bottomrule
    \end{tabular}
    }
    \end{center}
\vspace{-2em}
\end{table}
\section{Experiments}
\subsection{Setup}
\textbf{Data.}
Gigaspeech training set~\citep{GigaSpeech2021} is used as the training data, which contains 9k hours of audiobooks, podcasts, and YouTube videos at 16kHz audio sampling rate. Audio files that shorter than 2 seconds are dropped. For ablation studies, we use the masked reconstruction task, and a 1000-utterance random subset of Gigaspeech validation set as the testing utterances (detailed in \S\ref{sec:implment_detail}).
For speech editing evaluation, we use the proposed \dataname~dataset. For zero-shot TTS evaluation, we constructed a $250$ prompt-transcript paired dataset from LibriTTS~\citep{libritts} 
and the YouTube portion of the Gigaspeech test set, with half of the examples drawn from each dataset. The length of each voice prompt is kept as close as possible to 3 seconds long, with the constraint applied that we only cut the audio between complete words. The transcript is a concatenation of the transcript of the voice prompt and the target transcript. The target transcripts are chosen from different utterances spoken by the same speaker as the prompt, and range from $8$ to $40$ words in length. We only select utterances with a WER lower than 15\% by Whisper medium.en~\cite{Radford2022RobustSR}.

\textbf{Model.} 
Encodec~\citep{Defossez2022HighFN} is used as the speech tokenizer, which has $4$ RVQ codebooks each with vocabulary size of $2048$, and a codec framerate of 50Hz on 16kHz recordings. (see \S\ref{sec:implment_detail} for detailed config). To choose the number of spans to mask in training, we use a Poison($1$) distribution truncated to a minimum of $1$ and maximum of $3$. Span lengths are sampled from Uniform($1$, $600$)
i.e. the masked speech can be as long as $12$ seconds.
At each time step, the  embeddings of codes from different codebooks are summed~\citep{Wang2023NeuralCL}, then added by sinusoidal positional encoding~\citep{Vaswani2017AttentionIA}, before being fed to the transformer.
Text transcripts are phonemized based on the IPA phoneset using the toolkit provided by~\citet{Bernard2021}. Our main \modelname~model has $16$ transformers layer with hidden/FFN dimensions of $2048$/$8192$, and $12$ attention heads. The output of the last layers are fed to four separate $2$-layer MLP modules to get prediction logits. Our Main model has 830M parameters and codebook weight hyperparameters $\alpha$ is set to be $(5,1,0.5,0.1)$. Ablations on model sizes and codebook weights are shown in \S\ref{sec:ab}.

\textbf{Training and inference.}
The training of the Encodec model largely follows the setting in ~\citet{musicgen}, detailed in \S\ref{sec:implment_detail}.
To train \modelname, we used the ScaledAdam optimizer and Eden Scheduler proposed in~\citep{Yao2023ZipformerAF} with a base learning rate of $0.05$, batch size of 400k frames (i.e. 133.2 minutes), and total training step of 50k with gradient accumulation. The training of the 830M \modelname~model took about 2 weeks on $4$ NVIDIA A40 GPUs. More details can be found in \S\ref{sec:implment_detail}. We compare the performance of ScaledAdam and AdamW in \S\ref{sec:opt}.
For inference, we use Nucleus sampling~\citep{Holtzman2020The} with $p=0.8$ and a temperature of $1$ for all experiments. Due to the stochasticity of autoregressive generation, via manual inspection we found that while most of the time the model produces natural sounding speech, it sometimes produces excessively long silence or drags out certain sounds. We found that happens when the codec token generation gets stuck in a repeating loop. To resolve it, we use a simple heuristic: for each input utterance we generate several different output utterances and throw away the longest outputs. Specifically for speech editing, we run inference $10$ times with different margin parameters, stepping $\epsilon$ up from $0.05$ to $0.14$ in $0.01$ increments. The $4$ longest outputs are discarded, and then we randomly select one sample from the remaining 6 outputs. For zero-shot TTS, we reduce the probability of generating the same token in consecutive timesteps in proportion to how many times that token was consecutively generated in the immediately preceding timesteps. In addition, we generate $5$ samples with different random seeds, and select the shortest for TTS evaluation. 
The sample selection process is completely automatic and unsupervised (i.e. no human intervention or ASR scoring).

\textbf{Baselines.}
For speech editing, we compare \modelname~with the diffusion-based model FluentSpeech~\citep{Jiang2023FluentSpeechSA} which is the current open-source SotA model for speech editing. Since the original FluentSpeech model is trained on LibriTTS, for a fair comparison, 
we took the official GitHub \href{https://github.com/Zain-Jiang/Speech-Editing-Toolkit}{repo} and trained the model on Gigaspeech.
Please find more details in \S\ref{sec:implment_detail}. 
For zero-shot TTS, we compare our \modelname~with VALL-E~\cite{Wang2023NeuralCL}, XTTS v2~\citep{xttsv2}, YourTTS~\citep{Casanova2021YourTTSTZ}, and FluentSpeech. Since the original VALL-E is not open-sourced, we use the code from the popular open-source implementation by~\citet{valle-open}, and also trained the model on Gigaspeech. XTTS v2 is a popular commercial zero-shot TTS model\footnote{The GitHub \href{https://github.com/coqui-ai/TTS}{repo} hosting XTTS v2 has 26k stars by Jan 2024.} trained on a mixture of publicly available data and web-crawled data, although the exact data sources are unknown. YourTTS is trained on VCTK, LibriTTS, and also French and Portugese corpora. 
\begin{table}
    \caption{Effect of scaling model sizes and codebook re-weighting. Lower is better for all metrics.}\label{tab:ab_size_weight}
    \vspace{-0.5cm}
    \begin{center}
    \resizebox{\columnwidth}{!}{
        \begin{tabular}{llcccc}
            \toprule
            Params&Weights&WER &MCD &F0&Energy\\
            \midrule
            120M&(1,1,1,1) & 10.18&8.75 & 78.49 & 3.22  \\
            120M&(5,1,0.5,0.1) & 7.75& 8.31 & 87.74 & 3.54\\
            430M&(1,1,1,1)&7.87&8.22&70.05&3.17 \\
            430M&(5,1,0.5,0.1)&7.30&8.13&73.41&3.19 \\
            830M &(5,1,0.5,0.1)& 6.68&8.05 & 67.81 & 3.12 \\
            \bottomrule
        \end{tabular}
    }
    \end{center}
    \vspace{-0.8cm}
\end{table}

\begin{table*}[htb]
    \caption{Performance comparison on speech editing.}
    \vspace{-0.5em}
    \begin{center}
    \resizebox{0.99\textwidth}{!}{
    \begin{tabular}{lccccaccca}
        \toprule
        &&\multicolumn{4}{c}{Intelligibility MOS}&\multicolumn{4}{c}{Naturalness MOS}\\\cmidrule(lr){3-6} \cmidrule(lr){7-10}  
        Model&WER&LibriTTS&YouTube&Spotify&Total&LibriTTS&YouTube&Spotify&Total\\
        \midrule
          FluentSpeech & \textbf{4.5} & 3.89{\tiny$\pm$0.09} & 4.08{\tiny$\pm$0.08} & 3.95{\tiny$\pm$0.08} & 3.97{\tiny$\pm$0.05} & 3.42{\tiny$\pm$0.10} & 4.07{\tiny$\pm$0.10} & 3.93{\tiny$\pm$0.10} & 3.81{\tiny$\pm$0.06} \\
          \modelname & 6.1 & \textbf{4.05}{\tiny$\pm$0.08} & \textbf{4.14}{\tiny$\pm$0.07} & \textbf{4.12}{\tiny$\pm$0.07} & \textbf{4.11}{\tiny$\pm$0.05} & \textbf{3.68}{\tiny$\pm$0.10} & \textbf{4.25}{\tiny$\pm$0.09} & \textbf{4.16}{\tiny$\pm$0.08} & \textbf{4.03}{\tiny$\pm$0.05} \\
        \cmidrule(lr){1-10}
          Original & 5.4&4.22{\tiny$\pm$0.07}&  4.30{\tiny$\pm$0.07} & 4.16{\tiny$\pm$0.08} & 4.22{\tiny$\pm$0.05} & 3.84{\tiny$\pm$0.09} & 4.35{\tiny$\pm$0.08} & 4.29{\tiny$\pm$0.08} & 4.17{\tiny$\pm$0.05} \\
        \bottomrule
    \end{tabular}
    }
    \vspace{-0.6em}
    \end{center}
    \label{tab:se_main}
\end{table*}

\begin{table}
    \caption{\small Side-by-side naturalness comparison of \modelname~(\textsc{VCr}) v.s. Original (Orig.) and  FluentSpeech (FS).}\label{tab:se_sxs}
    \vspace{-0.8em}
    \begin{center}
    \resizebox{\columnwidth}{!}{
        \begin{tabular}{lccc}
            \toprule
            Comparison & \textsc{VCr} better & Tie & \textsc{VCr} worse \\
            \midrule
            \modelname~v. FS & 56.1\% & 19.7\%& 24.1\%  \\
            \modelname~v. Orig. & 40.3\% & 16.2\% & 43.6\% \\
            \bottomrule
        \end{tabular}
    }
    \vspace{-2em}
    \end{center}
\end{table}
\begin{figure*}[h]
    \centering
    \includegraphics[width=\textwidth]{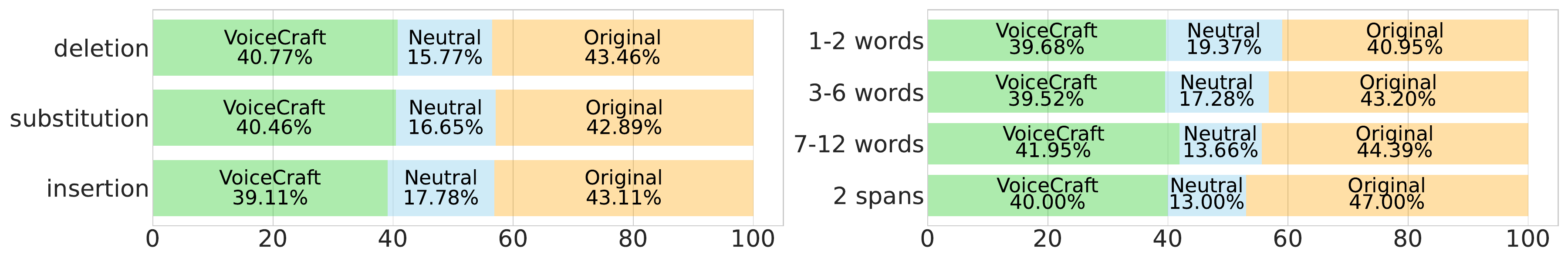}
    \vspace{-0.8cm}
    \caption{Breakdown of side-by-side human preference on naturalness comparing \modelname~edited speech and the original speech. Grouped by edit type (left) and edit span length (right).}
    \label{fig:se_sxs_vg_fs}
    \vspace{-0.5cm}
\end{figure*}
\begin{table*}[htb]
    \caption{On the zero-shot TTS task, comparing \modelname~with other models.}\label{tab:tts_main}
    \vspace{-1.2em}
    \begin{center}
      \resizebox{\textwidth}{!}{
    \begin{tabular}{lccccaccacca}
        \toprule
        &&&\multicolumn{3}{c}{Intelligibility MOS}&\multicolumn{3}{c}{Naturalness MOS}&\multicolumn{3}{c}{Speaker Similarity MOS}\\\cmidrule(lr){4-6} \cmidrule(lr){7-9} \cmidrule(lr){10-12} 
        Model&WER&SIM&Libri.&YouTube&Total&Libri.&YouTube&Total&Libri.&YouTube&Total\\
        \midrule
        YourTTS & 6.6&0.41 & 3.28{\tiny$\pm$0.11} & 3.01{\tiny$\pm$0.12} & 3.14{\tiny$\pm$0.08}&2.99{\tiny$\pm$0.12} & 2.59{\tiny$\pm$0.12} & 2.79{\tiny$\pm$0.08}&3.10{\tiny$\pm$0.12} & 2.49{\tiny$\pm$0.12} & 2.79{\tiny$\pm$0.09} \\
        FluentSpeech & \textbf{3.5}&0.47& 3.70{\tiny$\pm$0.11} & 3.65{\tiny$\pm$0.12} & 3.67{\tiny$\pm$0.08}&3.34{\tiny$\pm$0.11} & 3.43{\tiny$\pm$0.12} & 3.38{\tiny$\pm$0.08}&4.10{\tiny$\pm$0.09} & 3.92{\tiny$\pm$0.11} & 4.01{\tiny$\pm$0.07} \\
        VALL-E & 7.1&0.50 & 4.05{\tiny$\pm$0.09} & 3.94{\tiny$\pm$0.10} & 4.00{\tiny$\pm$0.07}&3.85{\tiny$\pm$0.10} & 3.86{\tiny$\pm$0.10} & 3.86{\tiny$\pm$0.07}&4.12{\tiny$\pm$0.10} & 4.02{\tiny$\pm$0.10} & 4.07{\tiny$\pm$0.07} \\
        XTTS v2 & 3.6&0.47  & 4.29{\tiny$\pm$0.09} & 3.97{\tiny$\pm$0.10} & 4.13{\tiny$\pm$0.07}&4.02{\tiny$\pm$0.09} & 3.90{\tiny$\pm$0.10} & 3.96{\tiny$\pm$0.07}&3.64{\tiny$\pm$0.12} & 3.25{\tiny$\pm$0.12} & 3.44{\tiny$\pm$0.08} \\
        \modelname & 4.5&\textbf{0.55} & \textbf{4.38}{\tiny$\pm$0.08} & \textbf{4.08}{\tiny$\pm$0.10} & \textbf{4.23}{\tiny$\pm$0.06}&\textbf{4.16}{\tiny$\pm$0.08} & \textbf{4.18}{\tiny$\pm$0.09} & \textbf{4.17}{\tiny$\pm$0.06}&\textbf{4.35}{\tiny$\pm$0.08} & \textbf{4.33}{\tiny$\pm$0.09} & \textbf{4.34}{\tiny$\pm$0.06} \\
        \cmidrule(lr){1-12}
        Ground Truth & 3.8&0.76& 4.37{\tiny$\pm$0.08} & 4.42{\tiny$\pm$0.08} & 4.39{\tiny$\pm$0.06}&4.32{\tiny$\pm$0.08} & 4.64{\tiny$\pm$0.06} & 4.48{\tiny$\pm$0.05}&4.26{\tiny$\pm$0.10} & 4.62{\tiny$\pm$0.08} & 4.44{\tiny$\pm$0.06} \\
        \bottomrule
    \end{tabular}
      }
    \end{center}
    \vspace{-1em}
\end{table*}
\textbf{Metrics.}
For ablation studies, since ground truth waveform is avaliable, in addition to WER (using Whisper medium.en as the ASR model), we use mel-ceptral distortion (MCD), F0 distance (F0) and energy distance (Energy). These are all objective metrics and their definitions are detailed in \S\ref{sec:implment_detail}. For speech editing and zero-shot TTS evaluation, we use a combination of objective and subjective metrics. For the objective metrics, we used WER and speaker similarity (SIM) following prior works\citep{Wang2023NeuralCL,Kharitonov2023SpeakRA}. 
SIM is calculated using the WavLM-TDCNN~\citep{Chen2021WavLMLS}. WER and SIM are calculated on all $310$ utterances in \dataname, and $250$ utterances in the zero-shot TTS dataset. For our subjective evaluation, we used the Amazon Mechanical Turk platform to conduct human listening tests. For speech editing, the outputs of our model on all $310$ utterances from \dataname~are evaluated by Turkers in terms of naturalness and intelligibility, and we use a $5$-point Likert scale where $1$ means poor and $5$ means excellent.
We also performed side-by-side A/B testing of \modelname's output against the original (non-edited) speech, as well as the edited speech produced by FluentSpeech. In both cases, Turkers were asked to determine which utterance sounds more natural. The Turkers can choose either one of the two, or indicate that they are equally natural. Each evaluation received $5$ ratings from 5 different Turkers. 
For zero-shot TTS, we randomly sampled $80$ utterances ($40$ from LibriTTS and $40$ from YouTube) from the original evaluation set, and asked Turkers to rate the naturalness, intelligibility, and speaker similarity of the generated speech to the reference prompt on a $5$-point Likert scale. Each evaluation received $10$ ratings. For all evaluations except the side-by-side comparison, Mean-Opinion-Score (MOS) with $95\%$ confidence interval are reported. For the side-by-side comparison, we report the percentage of the time one model is preferred over the other.
$64$ and $59$ Turkers participated in speech editing and TTS evaluation respectively. Please refer to \S\ref{sec:amt-instruction} for instructions and participants description.

\vspace{-.07in}
\subsection{Ablations}\label{sec:ab}
In table~\ref{tab:ab_size_weight}, we see that larger model sizes lead to better performance across all metrics
. In addition, we see a bigger gap between the bigger models, indicating the potential of further scaling model (and possibly training data) sizes. For the impact of codebook re-weighting, and we see that weighting earlier codebook heavier leads to better performance on intelligibility related metrics WER and MCD, while worse performance on prosody related metrics F0 and Energy\footnote{This can be regarded as a probing results that shows the properties of different codebooks in RVQ models. Since this is not the focus of our work, we do not conduct further experiment on this direction.}. We choose weight $(5,1,0.5,0.1)$ in our final 830M model because anecdotally, we found that \modelname~is stronger in prosody compared to intelligibility (similar properties about NCLMs are also found in~\citep{Jiang2023MegaTTSZT,Song2024ELLAVSN,Du2024VALLTDG})

\subsection{Speech Editing Results}\label{sec:se}
Table~\ref{tab:se_main} shows the results of speech editing evaluation in terms of WER, and human preference on intelligibility and naturalness. 
Our \modelname outperforms FluentSpeech on both intelligibility and naturalness MOS across different sources. Interestingly, FluentSpeech achieves a WER lower than the original recording ($4.5$ v.s. $5.4$), although its intelligibility MOS ($3.97$) is worse than both \modelname~($4.11$) and original recording ($4.22$). This suggests that ASR model and human judgement diverge on FluentSpeech's intelligibility. Anecdotally, we observe that FluentSpeech tends to produce dull and sometimes robotic speech~\footnote{please refer to our \href{https://jasonppy.github.io/VoiceCraft_web}{demo page} for examples}, and we hypothesize that this type of speech tends be more easily recognized by ASR, but is less intelligible to human ears. We notice this same phenomenon in our results on zero-shot TTS. 

Human listeners rate LibriTTS's naturalness lower than YouTube and Spotify on original speech (results on TTS is consistent with this). This suggests that to better evaluate speech synthesis in general, the research community should consider evaluating on other speech domains besides audiobooks as is commonly done.

Table~\ref{tab:se_sxs} presents side-by-side utterance naturalness comparison of \modelname~vs. FluentSpeech and \modelname~vs. the original, unedited speech. We observe that \modelname~is preferred over FluentSpeech $56.1\%$ of the time, with an additional $19.7\%$ of the time the two are tied. This means that $75.9\%$ of the time, human listeners' think \modelname~produces equal or more natural speech than FluentSpeech. 
Impressively, human listeners judge the edited speech produced by \modelname~to be equally or more natural than the original unedited speech $56.4\%$ of the time.
Fig.~\ref{fig:se_sxs_vg_gt} shows the breakdown of the side-by-side comparisons by edit type and edit span length. We see that compared to the original speech, \modelname~performs consistently well across different edit types, but human listeners think its outputs are slightly less natural with longer edit span(s).

\subsection{Zero-Shot TTS Results}\label{sec:tts}
Table~\ref{tab:tts_main} shows both objective and subjective evaluation on zero-shot TTS. We observe that \modelname~achieves the best results in both automatic speaker similarity metric SIM, and all human evaluation metrics. In particular, \modelname~is only slightly worse than ground truth in terms of intelligibility MOS ($4.23$ v.s. $4.39$), and speaker similarity MOS ($4.34$ v.s. $4.44$). The gap on naturalness is larger between \modelname~and ground truth ($4.17$ v.s. $4.48$), especially on YouTube utterances, which highlights the challenges of zero-shot TTS on noisy, in-the-wild data. The commercial model XTTS v2 comes second in terms of intelligibility and naturalness, and second to last on speaker similarity MOS. VALL-E achieves the second best on both automatic metric SIM and subjective metric speaker similarity MOS. Similarly to the speech editing results, ground truth YouTube utterances receive higher MOS scores than ground truth LibriTTS utterances in Table~\ref{tab:tts_main}, which again suggests that we should consider using more diverse data for future speech synthesis model evaluation.
Lastly, we again observe that FluentSpeech achieves lower WER than the ground truth, but receives much lower ratings in terms of intelligibility MOS from human listeners, indicating that WER could be misleading in evaluating intelligibility of speech synthesis systems\footnote{we also tried Whisper Large-v3, it gets WER of $4.1$ for ground truth, and $2.7$ for FluentSpeech.}.
\section{Conclusion}
We introduce a neural codec language model \modelname~that achieves state-of-the-art performance on speech editing and zero-shot TTS on in-the-wild data. The key lies in an innovative token rearrangement procedure which enables efficient and effective autoregressive codec generation with bidirectional context. In addition, we introduce a first-of-its-kind high quality, challenging, and realistic speech editing dataset \dataname, which we believe can reliably measure the practicality of speech editing models. 

\section{Limitations}
Given the advancement of made by \modelname, there are still limitations. First and foremost is the long silence and scratching sound that occasionally occur during generation. Although in this work, we overcome it with sampling multiple utterances and selecting the shorter ones, more elegant and efficient methods are needed. Another important aspect is AI safety, how can we watermark and detect synthesized speech? While watermarking and deepfake detection has attracted increasing attention in the research community, and remarkable progress has been made~\citep{Zhang2020OneClassLT,Yamagishi2021ASVspoof2A,Chen2023WavMarkWF,Roman2024ProactiveDO}, more advanced models such as \modelname~presents new opportunities and challenges to safety research. To facilitate speech synthesis and AI safety research, we fully open source our codebase and model weights.
\section{Ethical Implications}
The speech synthesis model \modelname~introduced in this work has both positive and negative implications.

On the positive side, \modelname~holds the promise of significant benefits across several domains. For individuals with speech impairments or who have lost the use of their voice, \modelname~could be transformative, enabling these individuals new ways to communicate with ease and clarity that were previously not possible. Content creators, whether they work in education, video production, or podcasting, could leverage \modelname~ to streamline their editing processes, making it easier to produce high-quality content without the need to re-record takes when they contain a small mistake. Furthermore, \modelname's ability to handle diverse accents without compromising on quality opens up new possibilities for creating synthetic data. This could, in turn, enhance speech recognition systems, such as Voicebox~\citep{Le2023VoiceboxTM}, by providing them with a richer and more varied dataset to learn from, thereby improving their accuracy and accessibility to users worldwide.

However, the potential negative impacts of \modelname~cannot be overlooked. One of the primary concerns is the model's potential to exacerbate existing biases, particularly those related to ethnicity. If not carefully monitored and corrected, these biases could lead to unequal performance across different groups, perpetuating and possibly even worsening existing disparities. Moreover, the ease with which voices can be cloned raises serious concerns about misuse, including impersonation and fraud. The ability to replicate someone's voice with only a few seconds of reference audio could be exploited to commit crimes or spread misinformation, posing significant ethical and security challenges. As such, while the benefits of \modelname~are clear and substantial, it is imperative to approach its deployment with caution, ensuring that measures are in place to mitigate these risks and protect against potential misuse.

Despite the concerns regarding impersonation and fraud associated with \modelname, there are compelling reasons to advocate for its release. Foremost among these is the opportunity it presents for the broader research community and technology developers to better understand and mitigate these negative impacts. By making these methods open source, we can catalyze the development of more robust countermeasures against the misuse of voice cloning technologies. This collaborative approach allows for the rapid identification of vulnerabilities and the exploration of innovative strategies to address them. Moreover, the authors of this work fully committed to advancing the field responsibly. We are actively working on pioneering deepfake detection and watermarking algorithms specifically designed for synthetic speech. By doing so, we not only acknowledge the potential risks associated with our technology but also take concrete steps to ensure its ethical use. This dual approach of open collaboration and dedicated research into safeguarding mechanisms reflects our commitment to fostering a technological ecosystem where the benefits of voice cloning can be realized while minimizing its potential for harm.
\section{Acknowledgements}
We thank Ziyue Jiang for providing guidance in running inference with and scaling up the FluentSpeech model. We thank students at \href{https://saltlab.cs.utexas.edu/}{SALT Lab} of UT Austin for helpful discussions. This work is supported in part by the National Science Foundation under Grant No. 2238605. 

\bibliography{custom}

\appendix

\newpage
\appendix
\section{Additional Experiments}\label{sec:app-res}
\subsection{Comparing ScaledAdam and AdamW}\label{sec:opt}
The hyperparameters settings of ScaledAdam can be found in table~\ref{tab:hyper}. For AdamW~\citep{Loshchilov2017DecoupledWD}, we tried 3 settings: 
\begin{itemize}
    \item setting1: peak learning rate: 1e-5, batch size: 3.3 min, update steps: 500k
    \item setting2: peak learning rate: 1e-4, batch size: 33.3 min (same as ScaledAdam), update steps: 80k
    \item setting3: peak learning rate: 1e-4, batch size: 3.3 min, update steps: 500k
\end{itemize}
For all settings, we use a linear scheduler which linear ramp up the learning rate to peak in first 8\% steps, and linearly decay it afterwards. We use the common default values for other hyperparameters, setting $\beta_1=0.9, \beta_2=0.999$, $\text{weight-decay} = 0.01$. All experiments are done on 4 A40 GPUs.
Results are shown in table~\ref{tab:ab_adam_scaledadam}.\footnote{We early stopped AdamW setting 2 at step 57k to save the compute, as it has already taken more time than the finished ScaledAdam job while the performance was worse.} We see that ScaledAdam achieves better performance in all metrics while using less compute. However we note that due to limitation in computational resources, we could not exhaust hyperparameter search for AdamW, therefore we do not over-generalize our finding here.

\begin{table*}
    \caption{ScaledAdam consistently outperforms AdamW across all metrics, while taking 10\% less time to train.}\label{tab:ab_adam_scaledadam}
    \begin{center}
        \begin{tabular}{lllcccc}
            \toprule
            Optimizer&Setting&Training Time&WER &MCD &F0&Energy \\
            \cmidrule(lr){1-7}
            AdamW&lr=1e-5, bsz=13.3min, steps=500k&262 hours & 16.45&8.91&196.15&5.94 \\
            AdamW&lr=1e-4, bsz=133.2min, steps=57k&273 hours & 10.77&8.45&117.38&4.91 \\
            AdamW&lr=1e-4, bsz=13.3min, steps=500k&262 hours &7.58&8.32&82.73&3.70 \\
            ScaledAdam&lr=3e-2, bsz=133.2min, steps=50k&\textbf{237 hours} &\textbf{7.30}&\textbf{8.13}&\textbf{73.41}&\textbf{3.19} \\
            \bottomrule
        \end{tabular}
    \end{center}
\end{table*}
\subsection{Breakdown of side-by-side human preference comparison.}\label{sec:vc_fs} 
\begin{figure*}
    \centering
    \includegraphics[width=\textwidth]{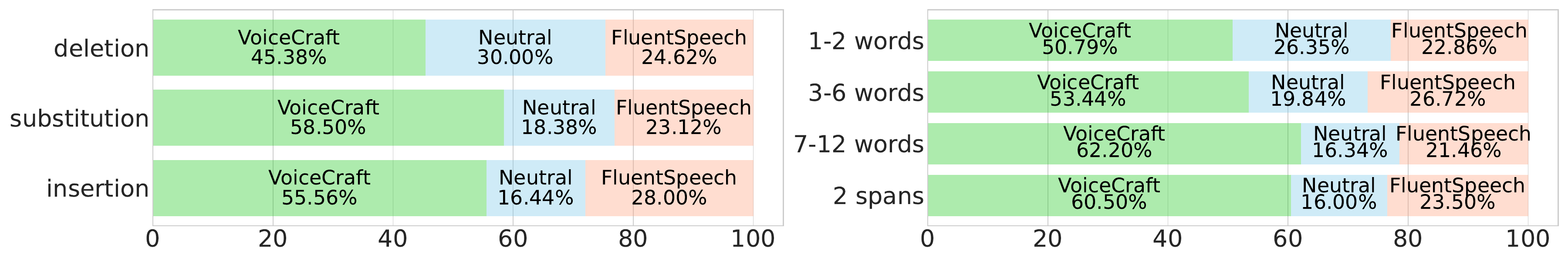}
    \vspace{-0.8cm}
    \caption{Breakdown of side-by-side human preference on naturalness comparing of \modelname~and FluentSpeech on speech editing. Grouped by edit type (left) and edit span length (right).}
    \label{fig:se_sxs_vg_gt}
\end{figure*}

The comparison breakdown between \modelname~and FluentSpeech is shown in figure~\ref{fig:se_sxs_vg_gt}. We see that \modelname~outperforms FluentSpeech across the board, especially for substitution edits and when the edit span length is long. 

\subsection{Spectrograms Comparison}\label{sec:spec}
\begin{figure*}
    \centering
    \includegraphics[width=1\linewidth]{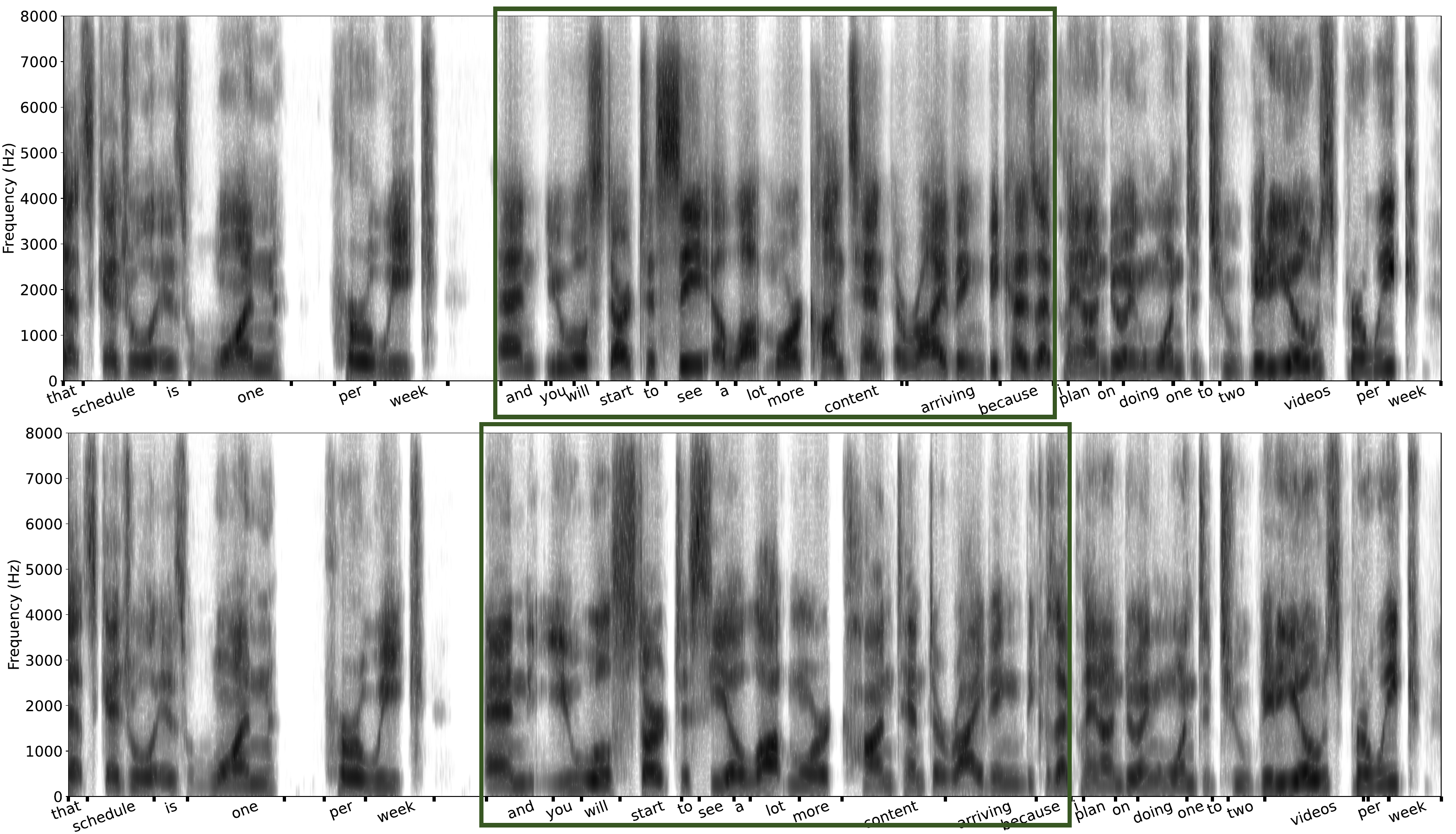}
    \caption{Upper: FluentSpeech; lower: \modelname}
    \label{fig:spec_content}
\end{figure*}
\begin{figure*}
    \centering
    \includegraphics[width=1\linewidth]{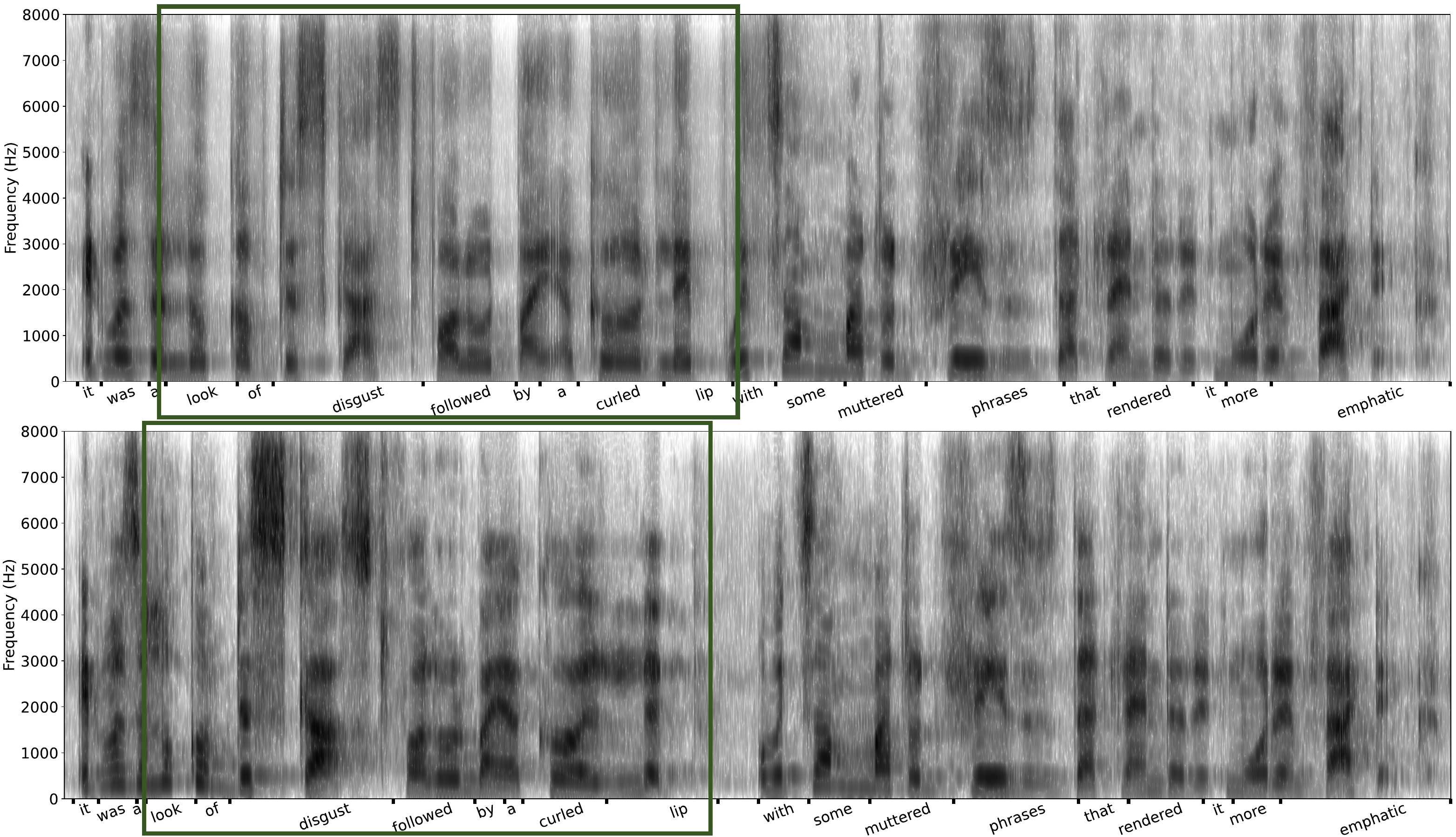}
    \caption{upper: FluentSpeech; lower: \modelname}
    \label{fig:spec_look}
\end{figure*}
\begin{figure*}
    \centering
    \includegraphics[width=1\linewidth]{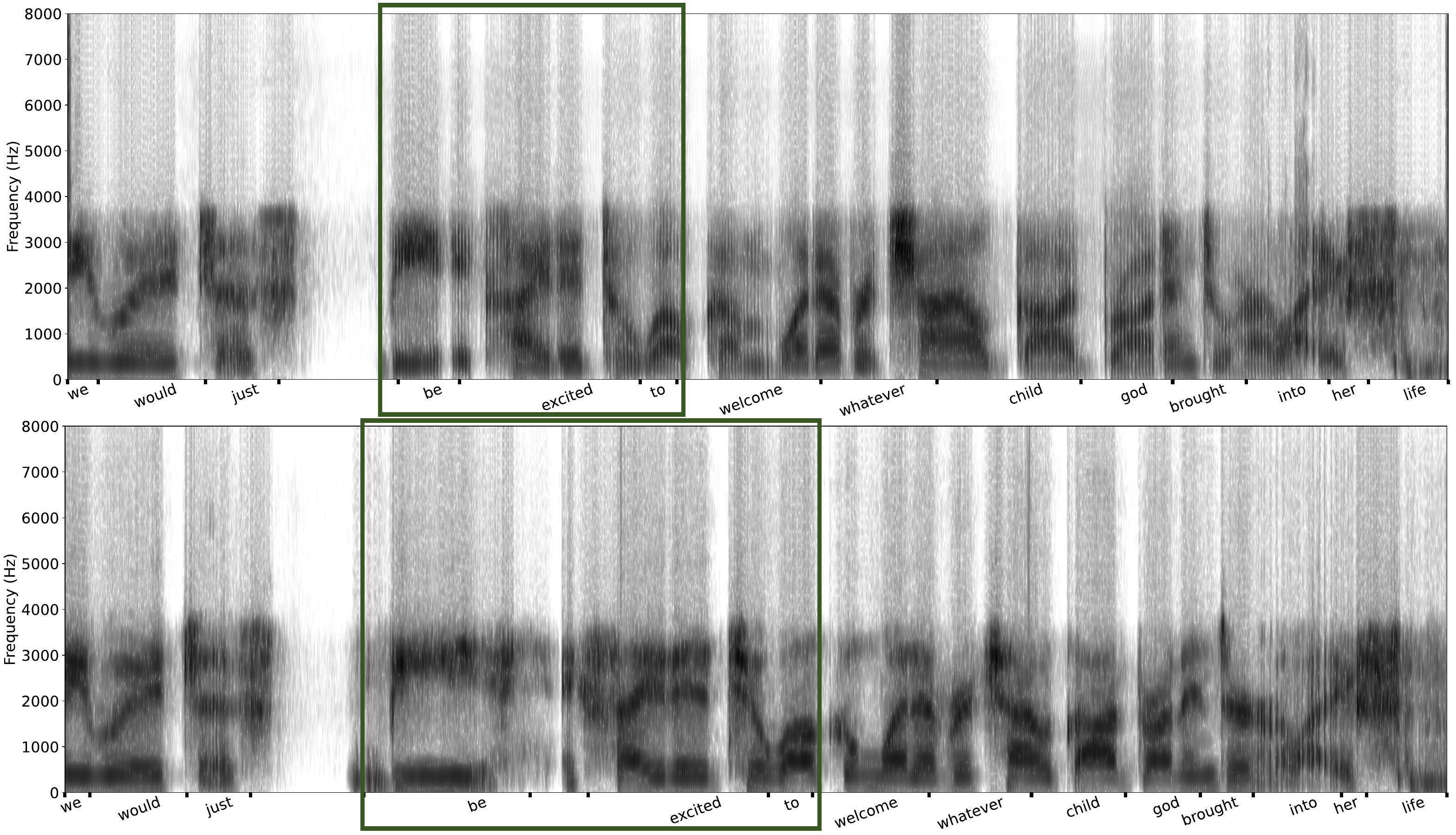}
    \caption{upper: FluentSpeech; lower: \modelname. Note that since the speech is recorded in very challenging condition, the word alignment is not very accurate. We see that for FluentSpeech's result, since the entire mel-spectrogram are passed to HiFi-GAN for resynthesis, even the unedited speech contains high frequency noise.}
    \label{fig:spec_child}
\end{figure*}
Spectrogram level comparison between FluentSpeech and \modelname are shown in figure~\ref{fig:spec_content},~\ref{fig:spec_look},~\ref{fig:spec_child} with the edited part marked in dark green rectangle. The three examples have increasing difficulty in terms of accents and recording conditions, in particular, the examples in figure~\ref{fig:spec_child} appears to be in low bandwidth transmission. In all 3 examples, we see that \modelname~is able to generated more detailed frequency patterns. The corresponding audio can be found in the demo page.
\section{Examples of the Speech Editing Dataset \dataname}\label{sec:app_data}
Examples of \dataname~are shown in table~\ref{tab:se_examples_app}.
\begin{table*}
    \caption{Examples of the speech editing dataset~\dataname.}\label{tab:se_examples_app}
    \centering
    \resizebox{\textwidth}{!}{%
    \begin{tabular}{L{0.13\textwidth}L{0.45\textwidth}L{0.45\textwidth}}
        \toprule
        Edit Types & Original & Edited \\
        \midrule
        substitution, substitution & See why it's extremely \textbf{valuable to it's kind of like} it's kind of like having a \textbf{wall hack} to watch a demo. & See why it's extremely \textbf{important right?} it's kind of like having a \textbf{rough time} to watch a demo. \\
        \midrule
        deletion & I wrote the title \textbf{of the course many years ago, ah,} when I created this course. & I wrote the \textbf{title when} I created this course. \\
        \midrule
        insertion & Fast cars, that had the nice \textbf{clothes, that} had the money, they was criminals. & Fast cars, that had the nice clothes, \textbf{that had expensive gold watches,} that had the money, they was criminals. \\
        \midrule
        substitution & When the CEO of blockbuster heard that, he promptly had \textbf{a kitchen sink} delivered to the netflix office, a fairly creative way of declaring war. & When the CEO of blockbuster heard that, he promptly had \textbf{five hundred pounds of glitter divided into five thousand manilla envelopes} delivered to the netflix office, a fairly creative way of declaring war. \\
        \midrule
        substitution & So if you've been following my story, you will remember that I said earlier \textbf{in this podcast that the} Grammy nominations came out. & So if you've been following my story, you will remember that I said earlier \textbf{that this week we had super exciting stuff to talk about because} Grammy nominations came out. \\
        \midrule
        insertion & No to the chemical pollution, air \textbf{pollution, and} the destruction of the environment caused by factories and the manufacturing industry. & No to the chemical pollution, air pollution, \textbf{no to the killing of plants and wildlife} and the destruction of the environment caused by factories and the manufacturing industry. \\
        \midrule
        substitution, substitution&because we can include so many other characters if we \textbf{just} expand the definitions to any \textbf{sword} wielder, who's a little spicy.&because we can include so many other participants if we \textbf{are brave enough to} expand the definitions to any \textbf{blade} wielder, who's a little spicy.\\
        \midrule
        insertion & So for more craziness now that French was \textbf{conquered we} have to join forces to Great Britain. & So for more craziness now that French was conquered \textbf{by the Germans,} we have to join forces to Great Britain. \\
        \midrule
        substitution&economic development remains one of the most \textbf{effective ways to increase the capacity} to adapt to climate change. &economic development remains one of the most \textbf{promising options that we have left on the table} to increase the capacity to adapt to climate change.\\
        \midrule
        insertion & And \textbf{we're at} this point. & And we're \textbf{all extremely excited} at this point. \\
        \midrule
        insertion & Steve also co-founded pixar animation studios. Which has revolutionized the film industry in it's short history \textbf{with brilliant} use of technology. & Steve also co-founded pixar animation studios. Which has revolutionized the film industry in it's short history with \textbf{films like toy story that showcase} brilliant use of technology. \\
        \midrule
        substitution, deletion & this is just so cozy \textbf{up} here, \textbf{and having that skylight is just lovely} isn't it. & this is just so cozy \textbf{and warm here, isn't} it. \\
        \midrule
        substitution & It was a \textbf{glance of inquiry, ending in a look of chagrin,} with some muttered phrases that rendered it more emphatic. & It was a \textbf{look of disgust followed by a curled lip,} with some muttered phrases that rendered it more emphatic. \\
        \midrule
        substitution & More of a base and infrastructure to \textbf{tell those stories rather than doing it} out of a out of a tent with solar power. & More of a base and infrastructure to \textbf{fight these battles instead of} out of a tent with solar power. \\
        \bottomrule
    \end{tabular}
    }
\end{table*}
\section{Implementational Details}\label{sec:implment_detail}

\textbf{The Encodec model.} The Encodec model we use has a stride of 320 samples, which means the codec framerate is 50Hz for recording of sample rate 16kHz. The base dimension is $64$, doubling at each of the $5$ convolutional layer in the encoder.
Following~\citep{musicgen}, we use the open-sourced audiocraft repo\footnote{Encodec training doc can be \href{https://github.com/facebookresearch/audiocraft/blob/main/docs/ENCODEC.md}{here}} for Encodec model training. $1$ second speech segments sampled from Gigaspeech over a total of $160$ epochs (320k steps) with a batch size of $240$. The model is trained with the Adam~\citep{Kingma2014AdamAM} with base learning rate of 3e-4.

\textbf{Eden Scheduler~\citep{Yao2023ZipformerAF}.} the scheduler adjust the learning rate $\alpha_t$ at step $t$ using the following formula:
\begin{align*}
    \alpha_t = &\alpha_{\text{base}} \cdot \left( \frac{t^2 + \alpha_{\text{step}}^2}{\alpha_{\text{step}}^2} \right)^{-0.25} \cdot \\ &\cdot\left( \frac{e^2 + \alpha_{\text{epoch}}^2}{\alpha_{\text{epoch}}^2} \right)^{-0.25} \cdot\\ &\cdot \text{linear}(\alpha_{\text{start}}, t_{\text{warmup}}, t).
\end{align*}
Where $\alpha_{\text{base}}$ base learning rate, $t$ is the step index, $e$ is the epoch index, and $\alpha_{\text{step}}$ and $\alpha_{\text{epoch}}$ controls the amount of data the model has seen before significantly reducing the learning rate. $\text{linear}(\alpha_{\text{start}}, t_{\text{warmup}},t)$ linearly increase the outcome from $\alpha_{\text{start}}$ to $1$ over $t_{\text{warmup}}$ steps, and stays at $1$. In our experiment, we set 
\begin{align*}
    \alpha_{\text{base}}=0.05, \alpha_{\text{step}}=3000, \alpha_{\text{epoch}}=4, \\
    \alpha_{\text{start}}=0.5,t_{\text{warmup}}=500
\end{align*}
Since our dataset is quite large, we use pseudo-epoch instead of the actual epoch, and $1$ pseudo-epoch is set to be $3000$ training steps. Note that the choice of these hyperparameters are inspired by~\citet{Yao2023ZipformerAF,valle-open}, and if computation resources permitted, a grid search might find better hyperparameters settings.

\textbf{Configuration in ablation studies.} Configuration of different models are shown in table~\ref{tab:hyper}. Note that we use base learning rate 3e-2 for 430M model instead of 5e-2 because the latter gave a NaN error. 
\begin{table*}
    \centering
    \begin{tabular}{lcccccc}
    \toprule
         Params&codebook dim& Trm hidden dim & FFN dim & Trm layers& Base LR & Update Steps \\
         \midrule
         830M& 2048 &2048&8192&16&5e-2&50k \\
         430M& 2048 &2048&8192&8&3e-2&50k \\
         120M& 1024 &1024&4196&8&5e-2&50k \\
         \bottomrule
    \end{tabular}
    \caption{Hyperparameters settings for the different model sizes. Trm stands for Transformer.}
    \label{tab:hyper}
\end{table*}

\textbf{Task and Data for ablation studies.} The evaluation task is masked reconstruction, where for each utterance, we randomly select a span of length $1$ to $15$ words to mask, and ask \modelname~to reconstruct the masked speech based on the transcript and unmasked speech. We use a $1000$-utterance random subset of the Gigaspeech validation set, which contains YouTube videos and podcast data. We ensure that each utterance in the subset has a WER lower than 15\% when decoded by Whisper medium.en~\citep{Radford2022RobustSR}.

\textbf{Metrics for ablation studies.} Since ground truth is available for masked reconstruction evaluation, in addition to WER (measured from Whisper medium.en's output), we also measure the mel-cepstral distortion (MCD)~\cite{Kubichek1993MelcepstralDM}, F0 distance (F0), and energy distance (Energy)
WER and MCD are better correlated with intelligibility of the speech, and F0 and Energy are better correlated with prosody similarity between the generated and ground truth.
MCD measures the difference of Mel Frequency Cepstrum Coefficients (MFCC) between generated and ground truth, defined as 
$$\text{MCD} = \frac{10}{\ln 10} \sqrt{\frac{1}{2} \sum_{i=1}^{L} (m^g_i - m^r_i)^2}$$
where $L$ is the order of MFCC, which we set to be $13$. $m^g_i$ is the $i$th MFCC of ground truth recording and $m^r_i$ is the $i$th MFCC of the generated. We use pymcd package~\footnote{https://github.com/chenqi008/pymcd}  for calculating MCD. For F0 estimation, we use the pYIN~\citep{Mauch2014PYINAF} algorithm implemented in librosa~\cite{McFee2015librosaAA} with minimal frequency 80hz and maximal frequency 600hz. For energy calculation, we use the root mean square of magnitude of spectrogram, which is extracted using short time Fourier transform with window length of $640$, hop size of $160$. Note that since generated speech might have a different length compared to ground truth, dynamic time wrapping is first applied to time aligned the extracted MFCC/F0/energy before calculating their euclidean distances.
For each model in the ablation study, we use $3$ different random seeds and report the averaged results.

\textbf{Scaling FluentSpeech.} The original FluentSpeech~\citep{Jiang2023FluentSpeechSA} is trained on LibriTTS, and we made our best effort in scaling it for a fair comparison. 
Taking guidance from the authors of FluentSpeech. 
We scale the batch size from $16$ utterances to $256$ utterances. Diffusion base hidden dimension from $320$ to $1024$, residual layers from $20$ layers to $30$ layers, residual channels from $256$ to $512$. The final model contains 330M parameters, which is roughly the same as the Voicebox model~\cite{Le2023VoiceboxTM}. The model was trained on Gigaspeech training set on 1 A40 GPU for 626k steps which took $10$ days. The HiFi-GAN vocoder is also retrained on Gigaspeech training set for 400k steps using hyperparameters used on Voicebox~\citep{Le2023VoiceboxTM} (they also use Hifi-GAN as vocoder to decode to 16kHz speech)

\textbf{Baselines for zero-shot TTS.}
For zero-shot TTS, we compare our \modelname~with VALL-E~\cite{Wang2023NeuralCL}, XTTS v2~\citep{xttsv2}, YourTTS~\citep{Casanova2021YourTTSTZ}, and FluentSpeech. Since the original VALL-E is not open-sourced, we use the code from the popular open-source implementation by~\citet{valle-open}, and also trained the model on Gigaspeech. Both the AR and NAR model are trained for 50k steps using the ScaledAdam optimizer and Eden scheduler, same as our \modelname. The commercial model XTTS v2 is composed of three modules, VQ-VAE~\citep{Oord2017NeuralDR} for speech tokenization, a GPT-2~\citep{Radford2019LanguageMA} model for speech token modeling and a customized HiFi-GAN~\citep{Kong2020HiFiGANGA} model for token to waveform generation. 
XTTS v2 is trained on a mixture of publicly available data and web-crawled data, but the exact data sources are unknown. 
YourTTS is a zero-shot TTS model built upon the adversarial VAE model VITS~\citep{Kim2021ConditionalVA}, with novel zero-shot multi-speaker and multilingual training. The model is trained on VCTK, LibriTTS, and also French and Portugese corpora. 
The FluentSpeech model we used for TTS is the same as in speech editing, as the model can be configured to do zero-shot TTS similar to Voicebox~\citep{Le2023VoiceboxTM}.

\textbf{Licenses of the speech corpora.} Licenses: LibriTTS: CC BY 4.0; Gigaspeech: Apache-2.0; Spotify Podcast dataset: CC BY 4.0.

\section{The Conditional Independence Assumption}\label{sec:independence}
To better explain the rational behind the conditional independent assumption in equation~\ref{eq:factorize}, we go back to sequence $Y$ produced by causal masking. The assumption we are making for equation~\ref{eq:factorize} to hold is equivalent to the assumption that given $W$ and $H_{s,t}$, $Y_{s,t,k}$ is independent of $I_{s,t,k}^{(1)}$ and $I_{s,t,k}^{(2)}$ defined as
\begin{align*}
    I_{s,t,k}^{(1)} &\triangleq (Y_{s,t+k-1,1}, Y_{s,t+k-2,2}, \cdots,Y_{s,t+1,k-1}) \\
    I_{s,t,k}^{(2)} &\triangleq (Y_{s,t-1,k+1}, Y_{s,t-2,k+2}, \cdots, Y_{s,t-K+k, K}) 
\end{align*}
We argue that this assumption is mild, because 1) $I_{s,t,k}^{(1)}$ are tokens from timestep after $t$ and therefore should have less impact on the distribution of $Y_{t,k}$ given past tokens $H_{s,t}$ ($H_{s,t}$ might also contain also future tokens in physical time if $Z_{s,t}$ is in the masked spans); 2) although $I_{s,t,k}^{(2)}$ are tokens from timestep before $t$, they are from codebooks that are later than codebook $k$ in the residual quantization chain, meaning that they model the residual left by codebook $k$ (at the corresponding timesteps). Given the fact that $\{Y_{s,t-1,k}, Y_{s,t-2,k+1}, \cdots, Y_{s,t-K+k,K-1}\}\subset H_{s,t}$\footnote{A weaker condition holds for the first K tokens in unmasked spans (which accounts for at most 0.08s of speech for our models), but we omit the discussion here for simplicity}, meaning that the ``fitted parts'' are given, and therefore the ``unfitted parts'' (which is the residual) should have miner impact on the distribution of $Y_{s,t,k}$. Empirically, MusicGen shows that a codec language model trained with the Delay Pattern enjoys the efficiency of the naive parallel pattern, while achieving similar modeling performance as completely flattened sequence.
\section{Instructions for human listening test}\label{sec:amt-instruction}
\begin{figure*}
    \centering
    \includegraphics[width=1\linewidth]{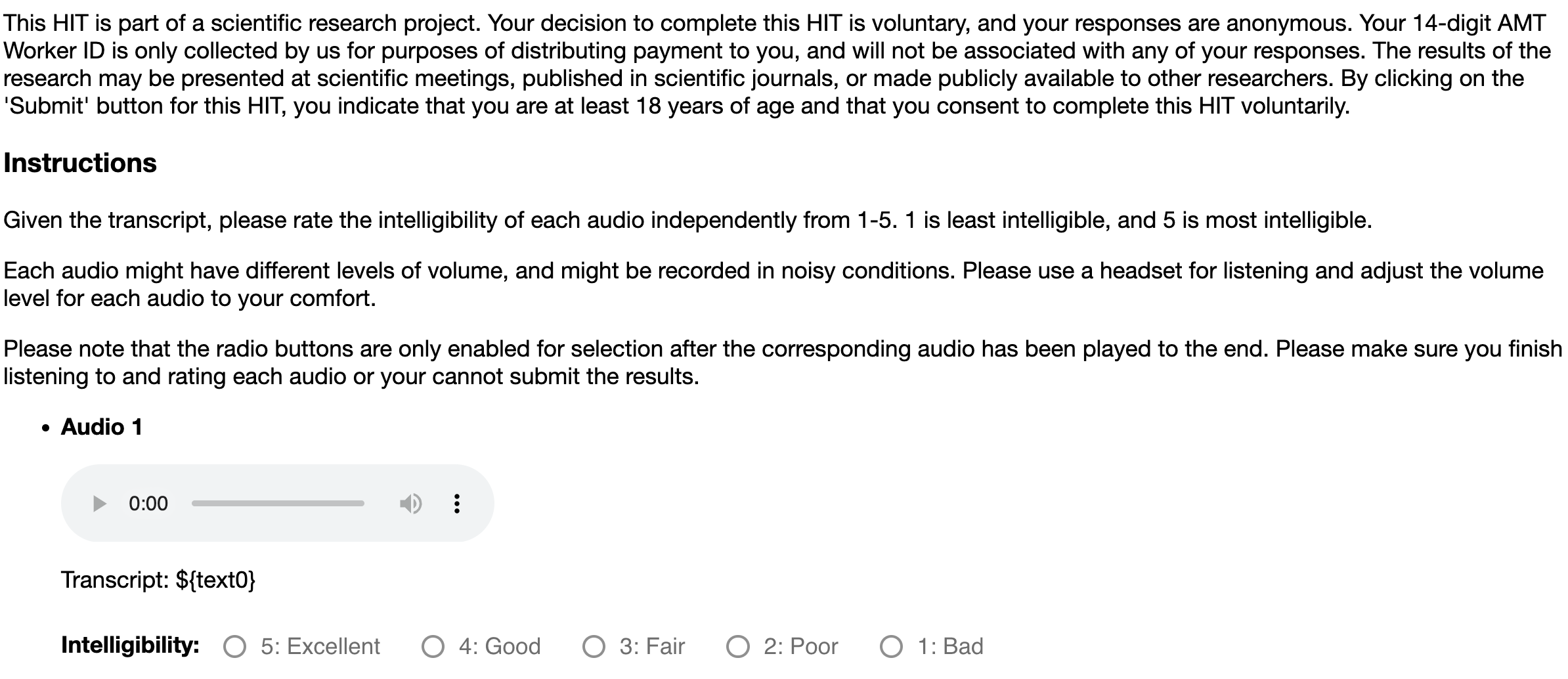}
    \caption{Instruction for speech editing-intelligibility preference. Each task contains 5 recordings. Since the first paragraph is also presented in all other tasks in the instruction page, we only show it in this screenshot.}\label{fig:se_in}
\end{figure*}

\begin{figure*}
    \centering
    \includegraphics[width=1\linewidth]{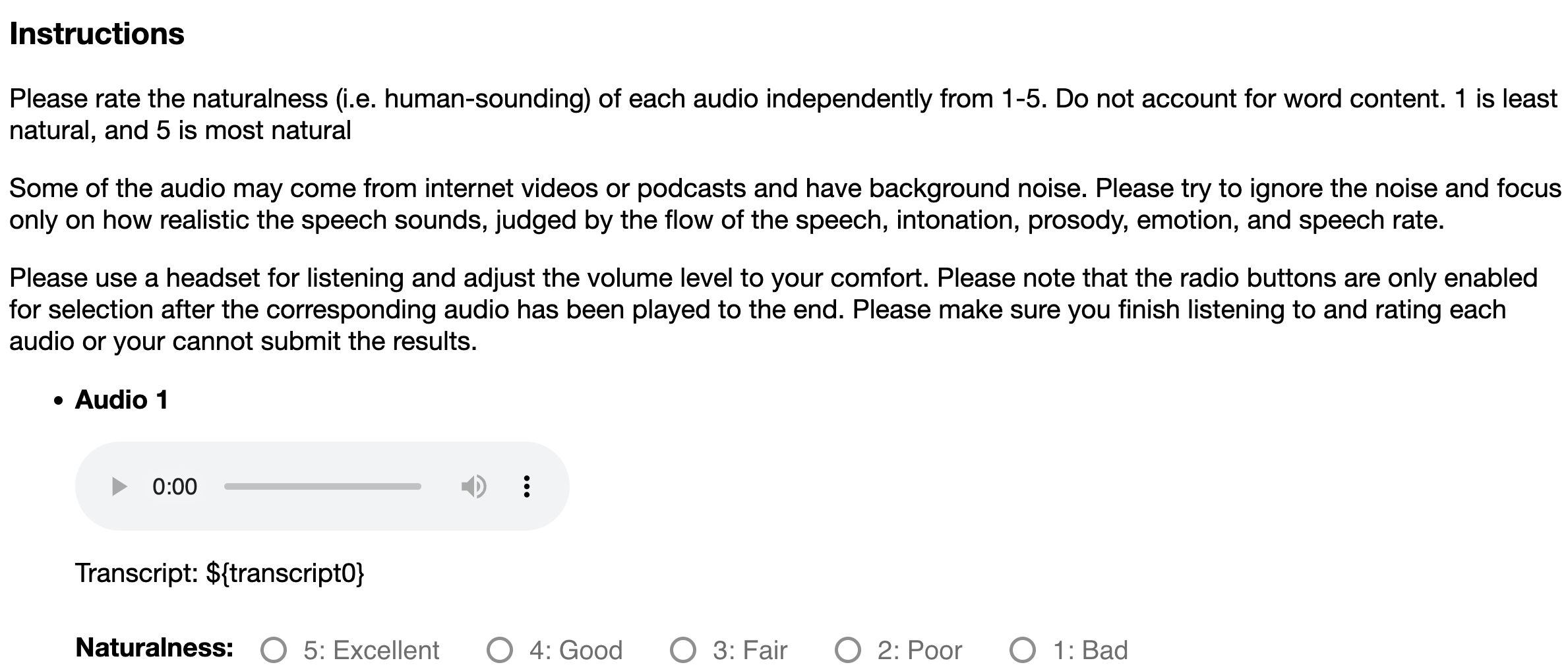}
    \caption{Instruction for speech editing-naturalness preference. Each task contains 5 recordings.}\label{fig:se_na}
\end{figure*}

\begin{figure*}
    \centering
    \includegraphics[width=1\linewidth]{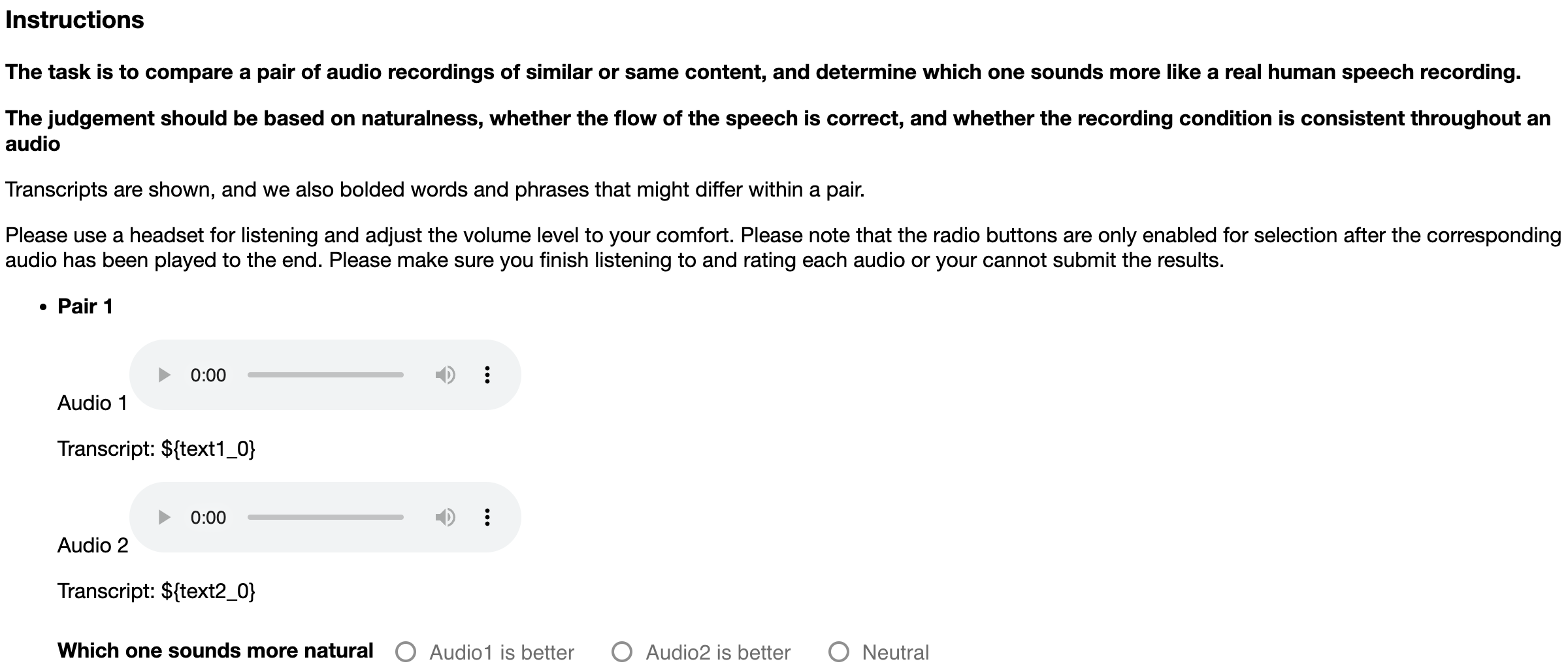}
    \caption{Instruction for speech editing, side-by-side naturalness preference. Each task contains 3 pairs of recordings.}\label{fig:se_si}
\end{figure*}

\begin{figure*}
    \centering
    \includegraphics[width=1\linewidth]{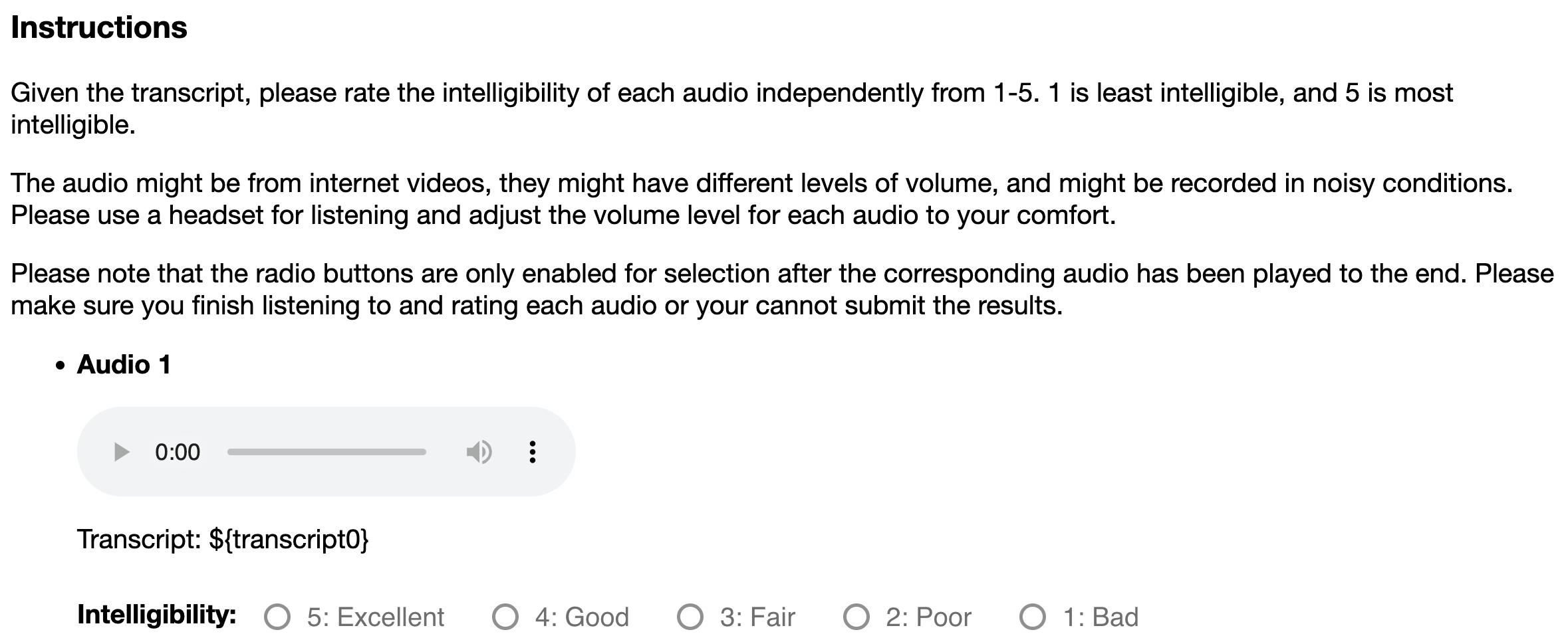}
    \caption{Instruction for zero-shot TTS, intelligibility preference. Each task contains 5 recordings.}\label{fig:tts_in}
\end{figure*}

\begin{figure*}
    \centering
    \includegraphics[width=1\linewidth]{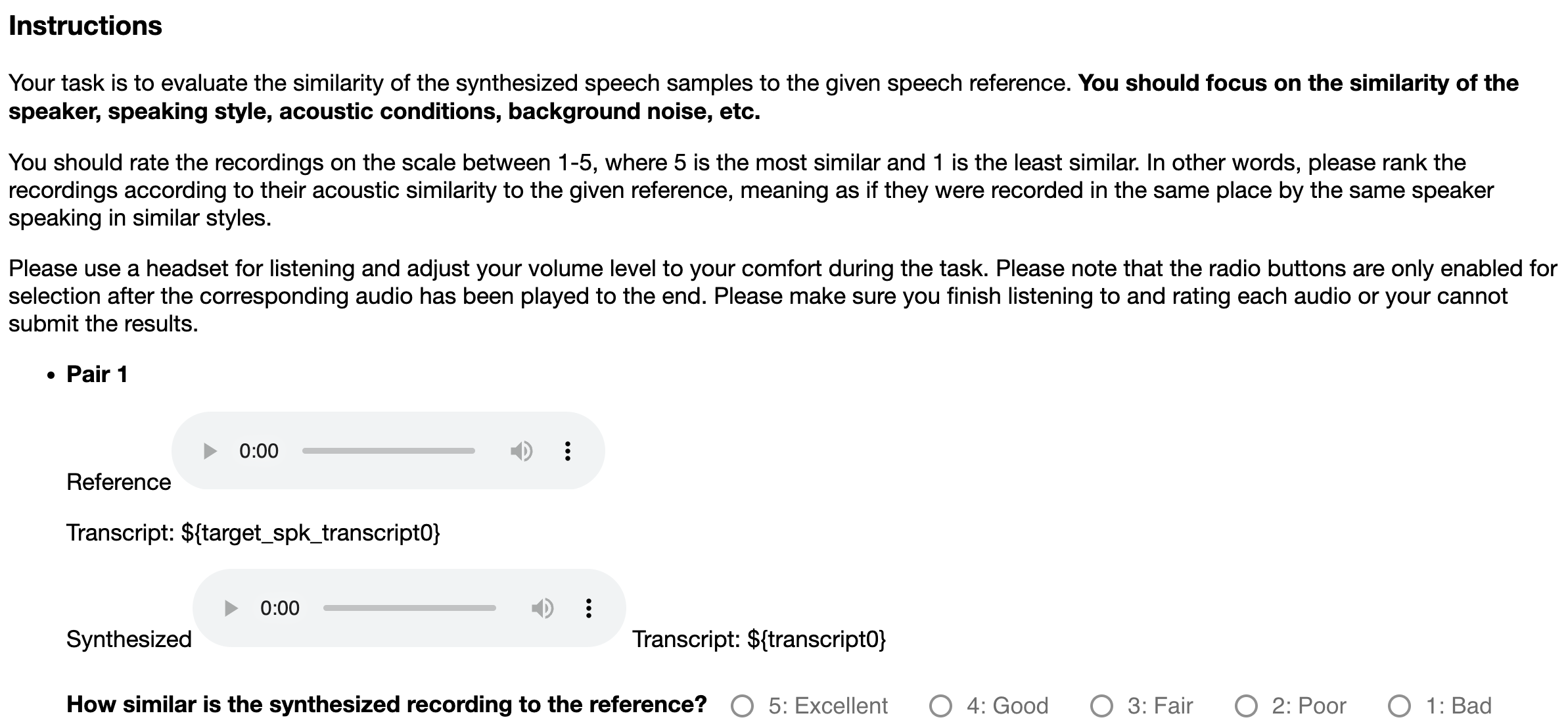}
    \caption{Instruction for zero-shot TTS, speaker similarity preference. Each task contains 3 pairs.}\label{fig:tts_spk}
\end{figure*}

\begin{figure*}
    \centering
    \includegraphics[width=1\linewidth]{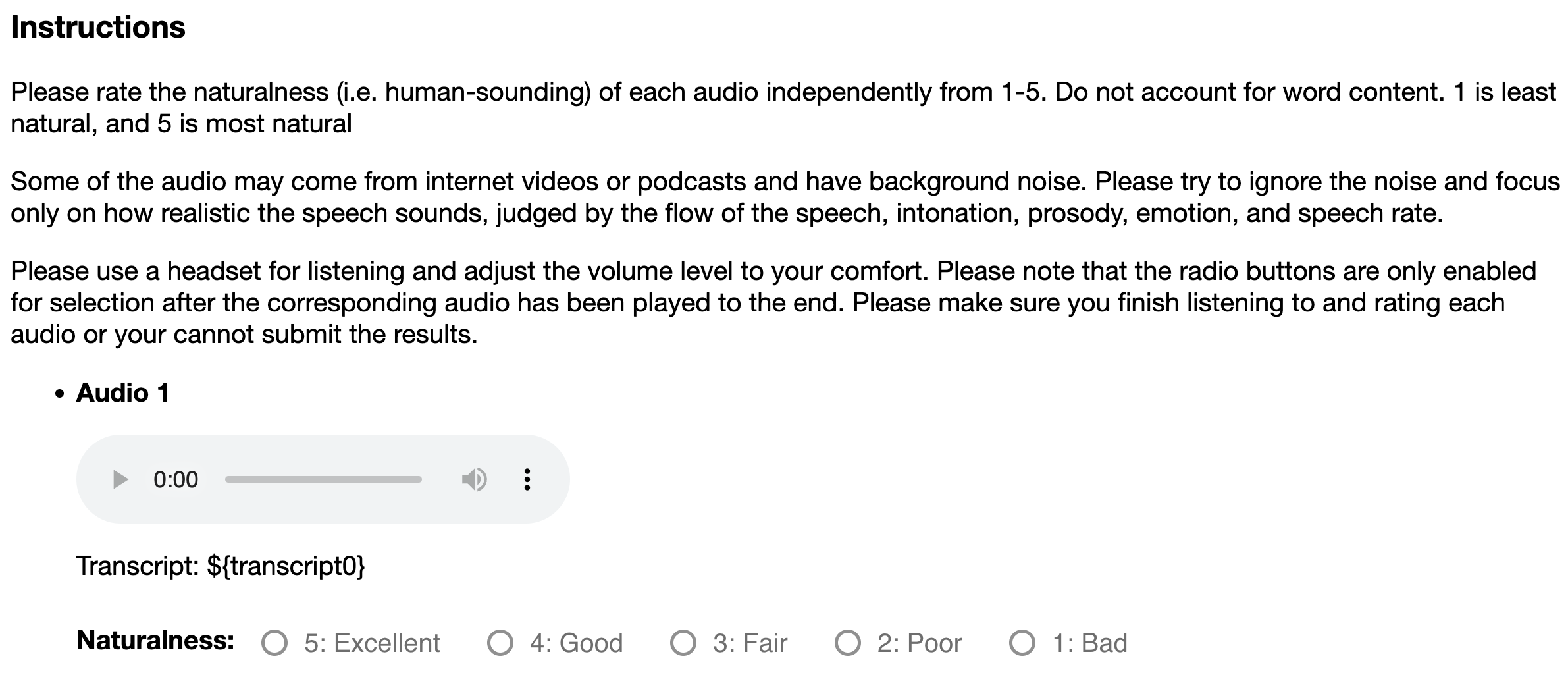}
    \caption{Instruction for zero-shot TTS, naturalness preference. Each task contains 5 recordings.}\label{fig:tts_na}
\end{figure*}

Screenshots of instructions for the human listening test we used on Amazon Mechanical Turk is shown in figure~\ref{fig:se_in} (speech editing - intelligibility), figure~\ref{fig:se_na}(speech editing - naturalness), figure~\ref{fig:se_si} (speech editing - side-by-side comparison), figure~\ref{fig:tts_in} (zero-shot TTS - intelligibility), figure~\ref{fig:tts_spk} (zero-shot TTS - speaker similarity), figure~\ref{fig:tts_na} (zero-shot TTS - naturalness). For speech editing evaluation, $64$ Turkers participated and we paid $474.3$ USD in total; for zero-shot TTS evaluation, $59$ Turkers participated and we paid $457.6$ USD. 
We only allow Turkers who are resident of the U.S. to do the tasks, and the goal is to increase the probability of Turkers being native English speakers. We acknowledge that this is a perfect approach and might need to bias in judgement, but since Amazon Mechanical Turk doesn't allow selection on native language, this is the best approach we could think of as a proxy to constraining the native language.

\end{document}